\documentclass[twocolumn]{autart}
\usepackage[dvipdfmx]{graphicx}
\usepackage{amsmath}
\usepackage{mathtools}
\usepackage{mathrsfs}
\usepackage{setspace}
\usepackage{bm}
\usepackage{comment}
\usepackage{amsfonts}
\usepackage{amssymb}
\usepackage{here}
\usepackage{natbib}
\usepackage{algorithm}
\usepackage{algorithmic}
\theoremstyle{definition}
\newtheorem{HMlem}{Lemma}

\newtheorem{HMthm}{Theorem}
\newtheorem{HMpro}{Proposition}
\newtheorem{HMassum}{Assumption}

\begin{document}
\begin{frontmatter}
\title{Separation and Estimation of Periodic/Aperiodic State\thanksref{footnoteinfo}}
\thanks[footnoteinfo]{
This work was supported by JST, ACT-X Grant Number JPMJAX200Q, Japan.
The material in this paper was not presented at any conference.\\
\hspace{1em}E-mail address: muramatsu@hiroshima-u.ac.jp, Tel.: +81-82-424-7575, Fax: +81-82-422-7193, Address: 1-4-1 Kagamiyama, Higashihiroshima, Hiroshima, 739-8527, Japan}
\author[Paestum]{Hisayoshi Muramatsu}\address[Paestum]{Graduate School of Advanced Science and Engineering, Hiroshima University, Higashihiroshima, 739-8527, Japan}
\begin{abstract}
Periodicity and aperiodicity can exist in a state simultaneously and typically become quasi-periodicity and quasi-aperiodicity in a dynamically changing state.
The quasi-periodic and quasi-aperiodic states existing in the periodic/aperiodic state mostly correspond to different phenomena and require different controls.
For separation control of these states, this paper defines the periodic/aperiodic, quasi-periodic, and quasi-aperiodic states to construct a periodic/aperiodic separation filter that separates the periodic/aperiodic state into the quasi-periodic and quasi-aperiodic states.
Based on these definitions, the linearity of periodic-pass and aperiodic-pass functions and the orthogonality of quasi-periodic and quasi-aperiodic state functions are proved.
Subsequently, the periodic/aperiodic separation filter composed of periodic-pass and aperiodic-pass filters that realize the periodic-pass and aperiodic-pass functions is designed and integrated with a Kalman filter for estimation of the quasi-periodic and quasi-aperiodic states.
\end{abstract}

\begin{keyword}
Periodic/Aperiodic separation filter,
Lifting,
Time delay,
Comb filter,
Kalman filter
\end{keyword}

\end{frontmatter}
\section{Introduction}
\begin{figure*}[t!]
	\begin{center}
		\includegraphics[width=0.85\hsize]{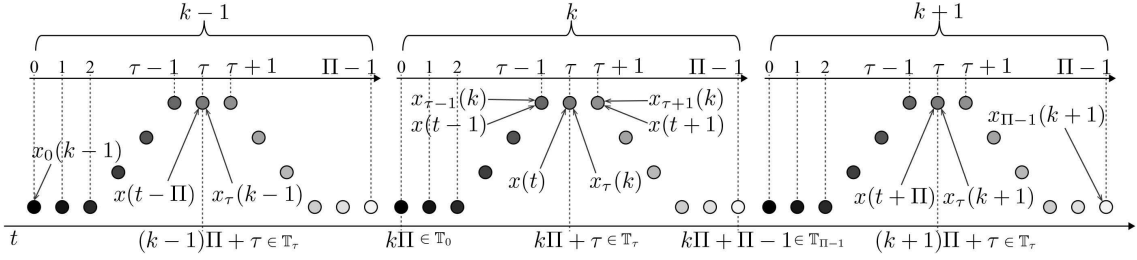}
	\caption{Relationship of $t$, $\Pi$, $k$, $\tau$, and $\mathbb{T}_\tau$ for the lifting function $L$.}\label{fig:def}
	\end{center}
\end{figure*}

Periodicity and aperiodicity are typical trends of states and systems; in particular, control strategies with periodicity have been widely studied.
Repetitive control proposed by \cite{1981_Inoue_RC,1988_Hara_RC} can realize precise periodic state control and has been developed to improve the precision (\cite{2006_Bristow_RC,2014_Chen_DOBbasedRC,2016_nagahara_RC}).
Additionally, \cite{2018_Muramatsu_APDOB,2016_Karttunen_DOBforP} studied the periodicity of a disturbance to estimate and compensate for a periodic disturbance, and \cite{1985_Bittanti_PS,1992_Bamieh_PS} investigated controls for periodic systems.
To address the periodicity, \cite{2000_Bittanti_PS,2010_Ebihara_PS,2014_Markovsky_PS,2018_Yang_PS} used a lifting technique to represent the periodic system as a time-invariant system.
Although many control strategies with periodicity have been proposed, a control strategy with both periodicity and aperiodicity is rare.
Using the lifting technique, \cite{2019_Muramatsu_PASF} proposed a periodic/aperiodic separation filter (PASF) that separates a periodic/aperiodic state into a quasi-periodic state and a quasi-aperiodic state; they used the PASF for separation control of the quasi-periodic and quasi-aperiodic states existing in the periodic/aperiodic state.
The PASF is similar to a comb filter, which rejects a periodic signal (\cite{2013_Sugiura_Comb,2017_Liu_Comb,2018_aslan}), but differs from the comb filter in addressing the quasi-periodic and quasi-aperiodic signals.
Nevertheless, the periodic/aperiodic separation control studies had three issues: the definitions of the periodic/aperiodic, quasi-periodic, and quasi-aperiodic states were qualitative; linearity and orthogonality were not proved; and variations of the PASF were not considered.
This paper addresses the aforementioned issues and makes the following contributions: it provides new definitions of the periodic/aperiodic, quasi-periodic, and quasi-aperiodic states; proves the linearity of periodic-pass and aperiodic-pass functions and orthogonality of quasi-periodic and quasi-aperiodic state functions; designs high-order infinite-impulse-response (IIR) and finite-impulse-response (FIR) realizations of the PASF; proposes a KF-PASF integrating the PASF and a Kalman filter (KF) (\cite{1960_kalman_KF,2012_Auger_KF}); and proves the unbiased estimation and equivalent sum of periodic-error and aperiodic-error covariances.
As for the significance of this study, the new definitions enable us to prove the linearity and orthogonality, which demonstrate a possibility of the independent separation control of the quasi-periodic and quasi-aperiodic states.
The proposed high-order IIR and FIR realization improves the separation performance of the PASF, and the proposed KF-PASF realizes the prediction and estimation of the states under noise.
Furthermore, the proposed realization and KF-PASF are expected to improve the precision of the periodic/aperiodic separation control for robots (\cite{2020_Muramatsu_PAMC}); improve the accuracy of the aperiodic anomaly detection (\cite{2021_Muramatsu_LEAK}); and eliminate harmonics noise similar to and better than the comb filters.

\section{Periodic/Aperiodic State} \label{sec:2}
\subsection{Quasi-Periodic and Quasi-Aperiodic States in Periodic/Aperiodic State}\label{sec:2-1}
Consider a state
\begin{align}
	x(t) \in \mathbb{R},\ t\in \mathbb{Z}.\notag
\end{align}
Using a period $\Pi\in \mathbb{Z}_{>0}$ of target periodicity, a lifting function $L:\mathbb{R}\to \mathbb{R}$ maps the state to a lifted state $x_{\tau}(k)$
\begin{align}
	&L(x(t)) \coloneqq x_{\tau}(k),\ x_{\tau}(k)\coloneqq x(t),\ k \in \mathbb{Z},\notag\\
	&\tau \coloneqq t\ \mathrm{mod}\ \Pi,\ \tau \in \{\zeta\in\mathbb{Z}|0\leq\zeta<\Pi\},\ k \coloneqq \frac{t-\tau}{\Pi},\notag
\end{align}
where the lifting function $L$ rewrites the domain with $t$ as the domain with $k$ and $\tau$.
The inverse lifting function $L^{-1}:\mathbb{R}\to \mathbb{R}$ is defined as
\begin{align}
	\label{eq:invlifting}
	&L^{-1}(x_{\tau}(k)) \coloneqq x(k\Pi+\tau)=x(t).
\end{align}
Fig.~\ref{fig:def} illustrates $t$, $\Pi$, $k$, and $\tau$ of the lifting function $L$.
In this paper, $x$ and $x_\tau$ are referred to as the state function and lifted state function, respectively.
The lifted state function $x_{\tau}$ belongs to a set of the lifted state functions $\mathbb{S}$, which is a set of all the Fourier transformable mappings from $\mathbb{Z}$ to $\mathbb{R}$
\begin{align}
	x_{\tau} \in \mathbb{S},\
	\mathbb{S} \coloneqq \mathbb{R}^\mathbb{Z} = \{(x_{\tau}(k))_{k\in \mathbb{Z}}|x_{\tau}(k)\in \mathbb{R}\}.\notag
\end{align}
The discrete-time Fourier transform $\mathcal{F}$ of $x_{\tau}(k)$ is
\begin{align}
	\label{eq:Fstate}
	&\mathcal{F}[x_{\tau}(k)] \coloneqq X_{\tau}(\omega) \coloneqq \displaystyle\sum_{k=-\infty}^{\infty}x_{\tau}(k)e^{-j\omega k},\\
	&\omega\in \{\omega\in\mathbb{R}|-\pi\leq\omega\leq\pi\},\notag
\end{align}
where $\omega=\tilde{\omega} \Pi T\ [\mathrm{rad/sample}]$, $\tilde{\omega}\ [\mathrm{rad/s}]$, and $T\ [\mathrm{s}]$ denote the normalized angular frequency, angular frequency, and sampling time, respectively.
Note that the sampling time of $t$ is $T$, and that of $k$ is $\Pi T$.
Subsequently, this paper defines a set of a zero function $\mathbb{S}_{\mathbb{O}}$, a set of lifted state functions that have quasi-periodicity $\mathbb{S}_\mathbb{P}$, and a set of lifted state functions that have quasi-aperiodicity $\mathbb{S}_\mathbb{A}$ as
\begin{subequations}
	\label{eq:DEF:sets}
\begin{align}
	\label{eq:DEF:S0}
	\mathbb{S}_{\mathbb{O}} &\coloneqq \{x_{\tau} \in \mathbb{S}|\forall\omega,\ X_{\tau}(\omega) = 0\},\\
	\label{eq:DEF:Sp}
	\mathbb{S}_\mathbb{P} &\coloneqq \{x_{\tau} \in \mathbb{S}|\exists\omega,\ X_{\tau}(\omega) \neq 0 \land|\omega|\leq\rho \},\\
	\label{eq:DEF:Sa}
	\mathbb{S}_\mathbb{A} &\coloneqq \{x_{\tau} \in \mathbb{S}|\exists\omega,\ X_{\tau}(\omega) \neq 0 \land \rho<|\omega| \},\\
	&\rho \in \{\rho\in \mathbb{R}|0\leq\rho\leq \pi\},
\end{align}
\end{subequations}
where the quasi-periodicity and quasi-aperiodicity are defined to be low-frequency waves and high-frequency waves of the lifted state function $x_{\tau}$, respectively.
This paper uses $\land$, $\lor$, $\Rightarrow$, and $\Leftrightarrow$ to denote the logical conjunction, disjunction, implication, and equivalence, respectively.
The variable $\rho\ [\mathrm{rad/sample}]$ denotes the normalized separation frequency, which is the boundary between the quasi-periodicity and quasi-aperiodicity.
The separation frequency $\tilde{\rho}\ [\mathrm{rad/s}]$ is given by
\begin{align}
	\label{eq:rho}
	\tilde{\rho}=\frac{\rho}{\Pi T}.
\end{align}
The whole set $\mathbb{S}$ is the union of the three sets
\begin{align}
	\mathbb{S}=\mathbb{S_O}\cup\mathbb{S_P}\cup\mathbb{S_A},\notag
\end{align}
and $\mathbb{S}_\mathbb{P}$ and $\mathbb{S}_\mathbb{A}$ do not contain $\mathbb{S_O}$
\begin{align}
	\mathbb{S_O} \not \subset (\mathbb{S_P}\cup\mathbb{S_A}).\notag
\end{align}
According to the sets, a lifted quasi-periodic-state function $x_{\tau\mathrm{p}}$ and lifted quasi-aperiodic-state function $x_{\tau\mathrm{a}}$ are defined by
\begin{subequations}
	\label{eq:DEF:x_tauap}
\begin{align}
	\label{eq:DEF:x_taup}
	x_{\tau \mathrm{p}}&\coloneqq x_{\tau}\ \mathrm{s.t.}\ x_{\tau}\in(\mathbb{S_O} \cup \mathbb{S_A})^\mathrm{c},\\
	\label{eq:DEF:x_taua}
	x_{\tau \mathrm{a}}&\coloneqq x_{\tau}\ \mathrm{s.t.}\ x_{\tau}\in(\mathbb{S_O} \cup \mathbb{S_P})^\mathrm{c},
\end{align}
\end{subequations}
and the quasi-periodic state $x_{\mathrm{p}}(t)$ and quasi-aperiodic state $x_{\mathrm{a}}(t)$ are defined by
\begin{align}
	x_{\mathrm{p}}(t) &\coloneqq L^{-1}(x_{\tau\mathrm{p}}(k)),\notag\\
	x_{\mathrm{a}}(t) &\coloneqq L^{-1}(x_{\tau\mathrm{a}}(k)).\notag
\end{align}
Note that the quasi-periodic state $x_{\mathrm{p}}(t)$ and quasi-aperiodic state $x_{\mathrm{a}}(t)$ are not low-frequency waves and high-frequency waves even though the lifted quasi-periodic state $x_{\tau\mathrm{p}}(k)$ and lifted quasi-aperiodic state $x_{\tau\mathrm{p}}(k)$ are low-frequency waves and high-frequency waves, respectively.
Furthermore, this paper defines the lifted periodic/aperiodic-state function:
\begin{align}
	x_{\tau \mathrm{pa}}&\coloneqq x_{\tau}\ \mathrm{s.t.}\ x_{\tau}\in\mathbb{S_P}\cap \mathbb{S_A}\notag
\end{align}
and the periodic/aperiodic state $x_{\mathrm{pa}}(t)$:
\begin{align}
	x_{\mathrm{pa}}(t) &\coloneqq L^{-1}(x_{\tau\mathrm{pa}}(k)),\notag
\end{align}
which is the sum of the quasi-periodic and quasi-aperiodic states
\begin{align}
	x_{\mathrm{pa}}(t)=x_{\mathrm{p}}(t)+x_{\mathrm{a}}(t).\notag
\end{align}
In summary, the state $x(t)$ varies as
\begin{align}
	\label{eq:x_variation}
	x(t)=\left\{
	\begin{array}{cl}
		x_{\mathrm{pa}}(t)&\mathrm{if}\ x_\tau \in \mathbb{S_P}\cap \mathbb{S_A}\\
		x_{\mathrm{p}}(t)&\mathrm{if}\ x_\tau\in (\mathbb{S_O} \cup \mathbb{S_A})^\mathrm{c}\\
		x_{\mathrm{a}}(t)&\mathrm{if}\ x_\tau\in (\mathbb{S_O} \cup \mathbb{S_P})^\mathrm{c}\\
		0&\mathrm{if}\ x_\tau \in \mathbb{S_O}
	\end{array}
	\right. .
\end{align}
Because the state function $x$ has $\Pi$ lifted state functions $x_0,\ \ldots,\ x_{\Pi-1}$, the domain $\mathbb{Z}$ of the state function $x$ has $\Pi$ corresponding subsets: $\mathbb{T}_0,\ \ldots,\ \mathbb{T}_{\Pi-1}$ defined by
\begin{align}
	\mathbb{T}_\tau\coloneqq\{t\in \mathbb{Z}|\forall k\in\mathbb{Z},\ t=k\Pi+\tau\},\notag
\end{align}
which is illustrated in Fig.~\ref{fig:def}.
In each domain $\mathbb{T}_\tau$, the state function $x$ is periodic/aperiodic if $x_\tau \in \mathbb{S_P}\cap \mathbb{S_A}$, quasi-periodic if $x_\tau \in (\mathbb{S_O} \cup \mathbb{S_A})^\mathrm{c}$, quasi-aperiodic if $x_\tau \in (\mathbb{S_O} \cup \mathbb{S_P})^\mathrm{c}$, or zero if $x_\tau \in \mathbb{S_O}$.

\begin{figure*}[t]
	\begin{center}
		\includegraphics[width=0.8\hsize]{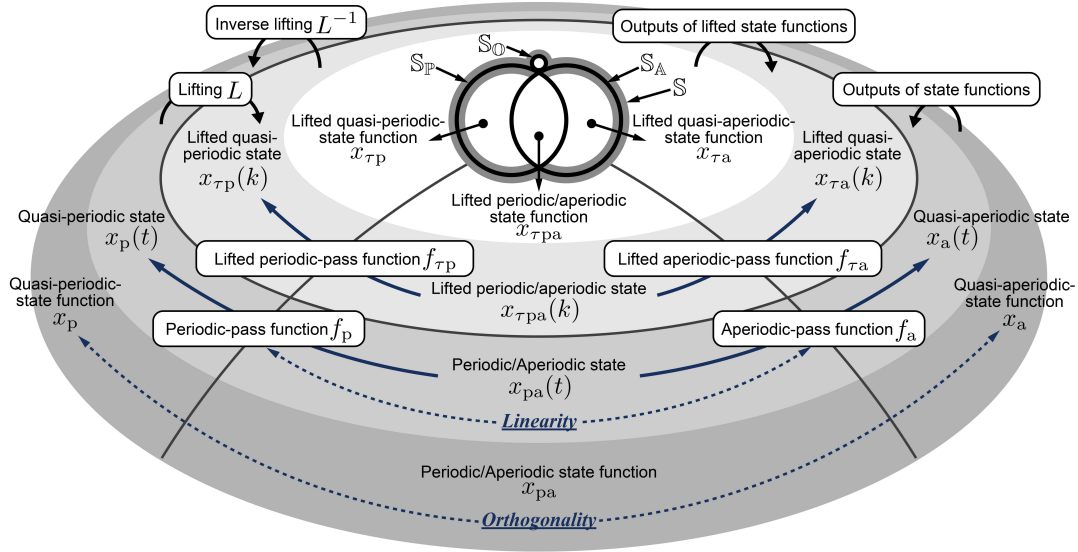}
	\caption{Relationship of the definitions, lifting, and separation for the sets, states, and state functions with quasi-periodicity and quasi-aperiodicity.}\label{fig:concept}
	\end{center}
\end{figure*}

\subsection{Linearity and Orthogonality} \label{sec:2-2}
Let $f_{\tau\mathrm{p}}:\mathbb{R}\to\mathbb{R}$ and $f_{\tau\mathrm{a}}:\mathbb{R}\to\mathbb{R}$ be the lifted periodic-pass function and lifted aperiodic-pass function, respectively
\begin{align}
	f_{\tau\mathrm{p}}(x_\tau(k))&\coloneqq\left\{
	\begin{array}{cl}
		x_{\tau\mathrm{p}}(k)&\mathrm{if}\ x_\tau \in \mathbb{S_P}\\
		0&\mathrm{if}\ x_\tau \in \mathbb{S_P}^{\mathrm{c}}
	\end{array}
	\right.,\notag\\
	f_{\tau\mathrm{a}}(x_\tau(k))&\coloneqq\left\{
	\begin{array}{cl}
		x_{\tau\mathrm{a}}(k)&\mathrm{if}\ x_\tau \in \mathbb{S_A}\\
		0&\mathrm{if}\ x_\tau \in \mathbb{S_A}^{\mathrm{c}}
	\end{array}
	\right. .\notag
\end{align}
Moreover, let $f_{\mathrm{p}}:\mathbb{R}\to\mathbb{R}$ and $f_{\mathrm{a}}:\mathbb{R}\to\mathbb{R}$ be the periodic-pass function and aperiodic-pass function, respectively
\begin{subequations}
	\label{eq:DEF:fpfa}
\begin{align}
	\label{eq:DEF:fp}
	f_{\mathrm{p}}(x(t))&\coloneqq\left\{
	\begin{array}{cl}
		x_{\mathrm{p}}(t)&\mathrm{if}\ x_\tau \in \mathbb{S_P}\\
		0&\mathrm{if}\ x_\tau \in \mathbb{S_P}^{\mathrm{c}}
	\end{array}
	\right.,\\
	\label{eq:DEF:fa}
	f_{\mathrm{a}}(x(t))&\coloneqq\left\{
	\begin{array}{cl}
		x_{\mathrm{a}}(t)&\mathrm{if}\ x_\tau \in \mathbb{S_A}\\
		0&\mathrm{if}\ x_\tau \in \mathbb{S_A}^{\mathrm{c}}
	\end{array}
	\right. ,
\end{align}
\end{subequations}
where
\begin{align}
	f_{\mathrm{p}}(x(t)) &= L^{-1}\Bigl(f_{\tau\mathrm{p}}\Bigl(L(x(t))\Bigr)\Bigr),\notag\\
	f_{\mathrm{a}}(x(t)) &= L^{-1}\Bigl(f_{\tau\mathrm{a}}\Bigl(L(x(t))\Bigr)\Bigr).\notag
\end{align}
Fig.~\ref{fig:concept} shows the relationship of the definitions, lifting, and separation for the sets, states, and state functions with quasi-periodicity and quasi-aperiodicity.
Preliminarily, this paper presents Lemma~\ref{LEM:FourierP/A}.
Then, for the states and functions, Theorems~\ref{THM:sum} and \ref{THM:product} demonstrate that the sets $\mathbb{S_A}^{\mathrm{c}}$ and $\mathbb{S_P}^{\mathrm{c}}$ are closed under addition and multiplication.
This implies that the quasi-periodicity and quasi-aperiodicity are retained or become zero after addition and multiplication.
Furthermore, Theorem~\ref{THM:linearity} demonstrates the linearity of the periodic-pass and aperiodic-pass functions.
Theorem~\ref{THM:orthogonality} and Proposition~\ref{PRO:pa} demonstrate the orthogonality of the quasi-periodic and quasi-aperiodic states and interference between the periodic-pass and aperiodic-pass functions, respectively.
\begin{HMlem}\label{LEM:FourierP/A}\end{HMlem}\vspace{-3mm}
\begin{align}
	&x_\tau \in \mathbb{S_A}^{\mathrm{c}} \Leftrightarrow
	X_{\tau}(\omega) =
	\left\{
	\begin{array}{cl}
		\displaystyle\sum_{k=-\infty}^{\infty}x_{\tau}(k)e^{-j\omega k}&\mathrm{if}\ |\omega|\leq\rho\\
		0&\mathrm{if}\ \rho<|\omega|
	\end{array}
	\right. ,\notag\\
	&x_\tau \in \mathbb{S_P}^{\mathrm{c}}\Leftrightarrow
	X_{\tau}(\omega) =
	\left\{
	\begin{array}{cl}
		0&\mathrm{if}\ |\omega|\leq\rho\\
		\displaystyle\sum_{k=-\infty}^{\infty}x_{\tau}(k)e^{-j\omega k}&\mathrm{if}\ \rho<|\omega|
	\end{array}
	\right. .\notag
\end{align}
$\bm{\mathrm{Proof.}}$
\eqref{eq:DEF:Sp} and \eqref{eq:DEF:Sa} give
\begin{align}
	x_\tau \in \mathbb{S_P} &\Leftrightarrow X_{\tau}(\omega) \neq 0\ \mathrm{if}\ |\omega|\leq\rho,\notag\\
	x_\tau \in \mathbb{S_A} &\Leftrightarrow X_{\tau}(\omega) \neq 0\ \mathrm{if}\ \rho<|\omega|,\notag
\end{align}
and their contrapositives:
\begin{align}
	x_\tau \in \mathbb{S_P}^{\mathrm{c}} &\Leftrightarrow X_{\tau}(\omega) = 0\ \mathrm{if}\ |\omega|\leq\rho,\notag\\
	x_\tau \in \mathbb{S_A}^{\mathrm{c}} &\Leftrightarrow X_{\tau}(\omega) = 0\ \mathrm{if}\ \rho<|\omega|.\notag
\end{align}
Using \eqref{eq:Fstate}, it is proved that
\begin{align}
	&x_\tau \in \mathbb{S_A}^{\mathrm{c}} \Leftrightarrow
	X_{\tau}(\omega) =
	\left\{
	\begin{array}{cl}
		\displaystyle\sum_{k=-\infty}^{\infty}x_{\tau}(k)e^{-j\omega k}&\mathrm{if}\ |\omega|\leq\rho\\
		0&\mathrm{if}\ \rho<|\omega|
	\end{array}
	\right. ,\notag\\
	&x_\tau \in \mathbb{S_P}^{\mathrm{c}}\Leftrightarrow
	X_{\tau}(\omega) =
	\left\{
	\begin{array}{cl}
		0&\mathrm{if}\ |\omega|\leq\rho\\
		\displaystyle\sum_{k=-\infty}^{\infty}x_{\tau}(k)e^{-j\omega k}&\mathrm{if}\ \rho<|\omega|
	\end{array}
	\right. .\notag
\end{align}
\mbox{ }\vspace{-10mm}
\begin{flushright} $\blacksquare$ \end{flushright}

\begin{HMthm}\label{THM:sum}\end{HMthm}\vspace{-3mm}
The sum of the quasi-periodic (quasi-aperiodic) states or zero is quasi-periodic (quasi-aperiodic) or zero.
\begin{align}
	&x_\tau,\ z_\tau\in \mathbb{S_A}^{\mathrm{c}}\Rightarrow y_\tau\in\mathbb{S_A}^{\mathrm{c}},\
	x_\tau,\ z_\tau\in \mathbb{S_P}^{\mathrm{c}}\Rightarrow y_\tau\in\mathbb{S_P}^{\mathrm{c}},\notag\\
	&y(t)\coloneqq x(t)+z(t),\ y_\tau(k)=x_\tau(k)+z_\tau(k).\notag
\end{align}
$\bm{\mathrm{Proof.}}$
Lemma~\ref{LEM:FourierP/A} gives
\begin{align}
	&x_\tau\in \mathbb{S_A}^{\mathrm{c}}\Leftrightarrow\notag\\
	&\mathcal{F}[x_{\tau}(k)] =
	\left\{
	\begin{array}{cl}
		\displaystyle\sum_{k=-\infty}^{\infty}x_{\tau}(k)e^{-j\omega k}&\mathrm{if}\ |\omega|\leq\rho\\
		0&\mathrm{if}\ \rho<|\omega|
	\end{array}
	\right.,\notag\\
	&x_\tau\in \mathbb{S_P}^{\mathrm{c}}\Leftrightarrow\notag\\
	&\mathcal{F}[x_{\tau}(k)] =
	\left\{
	\begin{array}{cl}
		0&\mathrm{if}\ |\omega|\leq\rho\\
		\displaystyle\sum_{k=-\infty}^{\infty}x_{\tau}(k)e^{-j\omega k}&\mathrm{if}\ \rho<|\omega|
	\end{array}
	\right.,\notag\\
	&z_\tau\in \mathbb{S_A}^{\mathrm{c}}\Leftrightarrow\notag\\
	&\mathcal{F}[z_{\tau}(k)] =
	\left\{
	\begin{array}{cl}
		\displaystyle\sum_{k=-\infty}^{\infty}z_{\tau}(k)e^{-j\omega k}&\mathrm{if}\ |\omega|\leq\rho\\
		0&\mathrm{if}\ \rho<|\omega|
	\end{array}
	\right. ,\notag\\
	&z_\tau\in \mathbb{S_P}^{\mathrm{c}}\Leftrightarrow\notag\\
	&\mathcal{F}[z_{\tau}(k)] =
	\left\{
	\begin{array}{cl}
		0&\mathrm{if}\ |\omega|\leq\rho\\
		\displaystyle\sum_{k=-\infty}^{\infty}z_{\tau}(k)e^{-j\omega k}&\mathrm{if}\ \rho<|\omega|
	\end{array}
	\right. .\notag
\end{align}
Using the linearity of the Fourier transform $\mathcal{F}[x_{\tau}(k) + z_{\tau}(k)] = \mathcal{F}[x_{\tau}(k)] + \mathcal{F}[z_{\tau}(k)]$, $\mathcal{F}[x_{\tau}(k) + z_{\tau}(k)]$ and $\mathcal{F}[x_{\tau}(k) + z_{\tau}(k)]$ are calculated as
\begin{align}
	&x_\tau,\ z_\tau\in \mathbb{S_A}^{\mathrm{c}}\Rightarrow\notag\\
	&\mathcal{F}[x_{\tau}(k) + z_{\tau}(k)] = \mathcal{F}[x_{\tau}(k)] + \mathcal{F}[z_{\tau}(k)]\notag\\
	&=\left\{
	\begin{array}{cl}
		\displaystyle\sum_{k=-\infty}^{\infty}[x_{\tau}(k) + z_{\tau}(k)]e^{-j\omega k}&\mathrm{if}\ |\omega|\leq\rho\\
		0&\mathrm{if}\ \rho<|\omega|
	\end{array}
	\right.,\notag\\
	&x_\tau,\ z_\tau\in \mathbb{S_P}^{\mathrm{c}}\Rightarrow\notag\\
	&\mathcal{F}[x_{\tau}(k) + z_{\tau}(k)] = \mathcal{F}[x_{\tau}(k)] + \mathcal{F}[z_{\tau}(k)]\notag\\
	&=\left\{
	\begin{array}{cl}
		0&\mathrm{if}\ |\omega|\leq\rho\\
		\displaystyle\sum_{k=-\infty}^{\infty}[x_{\tau}(k) + z_{\tau}(k)]e^{-j\omega k}&\mathrm{if}\ \rho<|\omega|
	\end{array}
	\right. ,\notag
\end{align}
and Lemma~\ref{LEM:FourierP/A} gives
\begin{align}
	&\left\{
	\begin{array}{cl}
		\displaystyle\sum_{k=-\infty}^{\infty}[x_{\tau}(k) + z_{\tau}(k)]e^{-j\omega k}&\mathrm{if}\ |\omega|\leq\rho\\
		0&\mathrm{if}\ \rho<|\omega|
	\end{array}
	\right.\notag\\
	&\Leftrightarrow y_\tau\in \mathbb{S_A}^{\mathrm{c}},\notag\\
	&\left\{
	\begin{array}{cl}
		0&\mathrm{if}\ |\omega|\leq\rho\\
		\displaystyle\sum_{k=-\infty}^{\infty}[x_{\tau}(k) + z_{\tau}(k)]e^{-j\omega k}&\mathrm{if}\ \rho<|\omega|
	\end{array}
	\right.\notag\\
	&\Leftrightarrow y_\tau\in \mathbb{S_P}^{\mathrm{c}}.\notag
\end{align}
They yield
\begin{align}
	&x_\tau,\ z_\tau\in \mathbb{S_A}^{\mathrm{c}}\Rightarrow y_\tau\in \mathbb{S_A}^{\mathrm{c}},\
	x_\tau,\ z_\tau\in \mathbb{S_P}^{\mathrm{c}}\Rightarrow y_\tau\in \mathbb{S_P}^{\mathrm{c}}.\notag
\end{align}
Thus, the sum of the quasi-periodic (quasi-aperiodic) states or zero is quasi-periodic (quasi-aperiodic) or zero.
This implies that the sum of the quasi-periodic (quasi-aperiodic) states is quasi-periodic (quasi-aperiodic) or zero, and it is trivial that the sum of the quasi-periodic (quasi-aperiodic) state and zero is quasi-periodic (quasi-aperiodic).
\mbox{ }\vspace{-6mm}
\begin{flushright} $\blacksquare$ \end{flushright}

\begin{HMthm}\label{THM:product}\end{HMthm}\vspace{-3mm}
The product of any value $a\in\mathbb{R}$ and the quasi-periodic (quasi-aperiodic) state or zero is quasi-periodic (quasi-aperiodic) or zero
\begin{align}
	&a \in \mathbb{R}\land x_\tau\in \mathbb{S_A}^{\mathrm{c}} \Rightarrow y_\tau \in \mathbb{S_A}^{\mathrm{c}},\notag\\
	&a \in \mathbb{R}\land x_\tau\in \mathbb{S_P}^{\mathrm{c}} \Rightarrow y_\tau \in \mathbb{S_P}^{\mathrm{c}},\notag\\
	&y(t)\coloneqq ax(t),\ y_\tau(k)=ax_\tau(k).\notag
\end{align}
$\bm{\mathrm{Proof.}}$
The Fourier transforms of the lifted quasi-periodic state $x_{\tau}(k)$ and lifted quasi-aperiodic state $x_{\tau}(k)$ multiplied by $a$ are
\begin{align}
	&a \in \mathbb{R}\land x_\tau \in \mathbb{S_A}^{\mathrm{c}}\Rightarrow
	a\mathcal{F}[x_{\tau}(k)] \notag\\
	&=a\left\{
	\begin{array}{cl}
		\displaystyle\sum_{k=-\infty}^{\infty}x_{\tau}(k)e^{-j\omega k}&\mathrm{if}\ |\omega|\leq\rho\\
		0&\mathrm{if}\ \rho<|\omega|
	\end{array}
	\right.,\notag\\
	&a \in \mathbb{R}\land x_\tau \in \mathbb{S_P}^{\mathrm{c}}\Rightarrow
	a\mathcal{F}[x_{\tau}(k)] \notag\\
	&=a\left\{
	\begin{array}{cl}
		0&\mathrm{if}\ |\omega|\leq\rho\\
		\displaystyle\sum_{k=-\infty}^{\infty}x_{\tau}(k)e^{-j\omega k}&\mathrm{if}\ \rho<|\omega|
	\end{array}
	\right.,\notag
\end{align}
which can be calculated using Lemma~\ref{LEM:FourierP/A} as
\begin{align}
	a \in \mathbb{R}\land x_\tau \in \mathbb{S_A}^{\mathrm{c}} \Rightarrow &
	\left\{
	\begin{array}{cl}
		\displaystyle\sum_{k=-\infty}^{\infty}ax_{\tau}(k)e^{-j\omega k}&\mathrm{if}\ |\omega|\leq\rho\\
		0&\mathrm{if}\ \rho<|\omega|
	\end{array}
	\right.\notag\\
	&\hspace{1em}\Leftrightarrow y_\tau\in \mathbb{S_A}^{\mathrm{c}},\notag\\
	a \in \mathbb{R}\land x_\tau \in \mathbb{S_P}^{\mathrm{c}} \Rightarrow &
	\left\{
	\begin{array}{cl}
		0&\mathrm{if}\ |\omega|\leq\rho\\
		\displaystyle\sum_{k=-\infty}^{\infty}ax_{\tau}(k)e^{-j\omega k}&\mathrm{if}\ \rho<|\omega|
	\end{array}
	\right.\notag\\
	&\hspace{1em}\Leftrightarrow y_\tau\in \mathbb{S_P}^{\mathrm{c}}.\notag
\end{align}
Therefore, the product of any value $a\in\mathbb{R}$ and the quasi-periodic (quasi-aperiodic) state or zero is quasi-periodic (quasi-aperiodic) or zero
\begin{align}
	a \in \mathbb{R}\land x_\tau\in \mathbb{S_A}^{\mathrm{c}} &\Rightarrow y_\tau \in \mathbb{S_A}^{\mathrm{c}},\notag\\
	a \in \mathbb{R}\land x_\tau\in \mathbb{S_P}^{\mathrm{c}} &\Rightarrow y_\tau \in \mathbb{S_P}^{\mathrm{c}}.\notag
\end{align}
\mbox{ }\vspace{-10mm}
\begin{flushright} $\blacksquare$ \end{flushright}

\begin{HMthm}\label{THM:linearity}\end{HMthm}\vspace{-3mm}
The periodic-pass function $f_{\mathrm{p}}(x(t))$ and aperiodic-pass function $f_{\mathrm{a}}(x(t))$ are linear
\begin{align}
	f_{\mathrm{p}}(x(t)+z(t))&=f_{\mathrm{p}}(x(t))+f_{\mathrm{p}}(z(t)),\ \forall x(t),\ z(t) \in \mathbb{R},\notag\\
	f_{\mathrm{a}}(x(t)+z(t))&=f_{\mathrm{a}}(x(t))+f_{\mathrm{p}}(z(t)),\ \forall x(t),\ z(t) \in \mathbb{R},\notag\\
	f_{\mathrm{p}}(ax(t))&=af_{\mathrm{p}}(x(t)),\ \forall x(t),\ z(t) \in \mathbb{R},\notag\\
	f_{\mathrm{a}}(ax(t))&=af_{\mathrm{a}}(x(t)),\ \forall x(t),\ z(t) \in \mathbb{R}.\notag
\end{align}
$\bm{\mathrm{Proof.}}$
The sum of the states $x(t)$ and $z(t)$ can be expressed by \eqref{eq:x_variation} as
\begin{align}
	&x(t)+z(t)=y_1(t)+y_2(t)+y_3(t)+y_4(t),\notag\\
	&y_1(t)\coloneqq\left\{
	\begin{array}{cl}
		x_{\mathrm{p}}(t)&\mathrm{if}\ x_\tau \in \mathbb{S_P}\\
		0&\mathrm{if}\ x_\tau \in \mathbb{S_P}^{\mathrm{c}}
	\end{array}
	\right.,\
	y_2(t)\coloneqq\left\{
	\begin{array}{cl}
		x_{\mathrm{a}}(t)&\mathrm{if}\ x_\tau \in \mathbb{S_A}\\
		0&\mathrm{if}\ x_\tau \in \mathbb{S_A}^{\mathrm{c}}
	\end{array}
	\right.,\notag\\
	&y_3(t)\coloneqq\left\{
	\begin{array}{cl}
		z_{\mathrm{p}}(t)&\mathrm{if}\ z_\tau \in \mathbb{S_P}\\
		0&\mathrm{if}\ z_\tau \in \mathbb{S_P}^{\mathrm{c}}
	\end{array}
	\right.,\
	y_4(t)\coloneqq\left\{
	\begin{array}{cl}
		z_{\mathrm{a}}(t)&\mathrm{if}\ z_\tau \in \mathbb{S_A}\\
		0&\mathrm{if}\ z_\tau \in \mathbb{S_A}^{\mathrm{c}}
	\end{array}
	\right. .\notag
\end{align}
Hence,
\begin{align}
	y_{1\tau} \in \mathbb{S_A}^{\mathrm{c}},\
	y_{2\tau} \in \mathbb{S_P}^{\mathrm{c}},\
	y_{3\tau} \in \mathbb{S_A}^{\mathrm{c}},\
	y_{4\tau} \in \mathbb{S_P}^{\mathrm{c}},\notag
\end{align}
and Theorem~\ref{THM:sum} gives
\begin{align}
	&(L(y_1(t)+y_3(t)))_{k\in \mathbb{Z}}\in \mathbb{S_A}^{\mathrm{c}},\
	(L(y_2(t)+y_4(t)))_{k\in \mathbb{Z}}\in \mathbb{S_P}^{\mathrm{c}}.\notag
\end{align}
Then, the periodic-pass and aperiodic-pass functions $f_{\mathrm{p}}$ and $f_{\mathrm{a}}$ output
\begin{align}
	f_{\mathrm{p}}(x(t)+z(t))&=y_1(t)+y_3(t),\notag\\
	f_{\mathrm{a}}(x(t)+z(t))&=y_2(t)+y_4(t).\notag
\end{align}
Using
\begin{align}
	f_{\mathrm{p}}(x(t))&=y_1(t),\
	f_{\mathrm{a}}(x(t))=y_2(t),\notag\\
	f_{\mathrm{p}}(z(t))&=y_3(t),\
	f_{\mathrm{a}}(z(t))=y_4(t),\notag
\end{align}
based on \eqref{eq:DEF:fp} and \eqref{eq:DEF:fa}, the additivity is obtained as
\begin{subequations}
	\label{eq:additivity}
\begin{align}
	\label{eq:additivity1}
	f_{\mathrm{p}}(x(t)+z(t))&=f_{\mathrm{p}}(x(t))+f_{\mathrm{p}}(z(t)),\\
	\label{eq:additivity2}
	f_{\mathrm{a}}(x(t)+z(t))&=f_{\mathrm{a}}(x(t))+f_{\mathrm{a}}(z(t)).
\end{align}
\end{subequations}
Next, according to \eqref{eq:x_variation},
\begin{align}
	&ax(t)=ay_1(t)+ay_2(t).\notag
\end{align}
Theorem~\ref{THM:product} gives
\begin{align}
	(L(ay_1(t)))_{k\in \mathbb{Z}} \in \mathbb{S_A}^{\mathrm{c}},\
	(L(ay_2(t)))_{k\in \mathbb{Z}} \in \mathbb{S_P}^{\mathrm{c}};\notag
\end{align}
hence,
\begin{align}
	f_{\mathrm{p}}(ax(t))&=ay_1(t),\
	f_{\mathrm{a}}(ax(t))=ay_2(t).\notag
\end{align}
Then, using
\begin{align}
	f_{\mathrm{p}}(x(t))&=y_1(t),\
	f_{\mathrm{a}}(x(t))=y_2(t),\notag
\end{align}
based on \eqref{eq:DEF:fpfa}, the homogeneity is obtained as
\begin{subequations}
	\label{eq:homogeneity}
\begin{align}
	\label{eq:homogeneity1}
	f_{\mathrm{p}}(ax(t))&=af_{\mathrm{p}}(x(t)),\\
	\label{eq:homogeneity2}
	f_{\mathrm{a}}(ax(t))&=af_{\mathrm{a}}(x(t)).
\end{align}
\end{subequations}
\eqref{eq:additivity} and \eqref{eq:homogeneity} prove that the periodic-pass function $f_{\mathrm{p}}(x(t))$ and aperiodic-pass function $f_{\mathrm{a}}(x(t))$ are linear.
\mbox{ }\vspace{-6mm}
\begin{flushright} $\blacksquare$ \end{flushright}

\begin{HMthm}\label{THM:orthogonality}\end{HMthm}\vspace{-3mm}
The quasi-periodic-state function $x_{\mathrm{p}}$ and quasi-aperiodic-state function $x_{\mathrm{a}}$ are orthogonal to each other
\begin{align}
	\sum_{t=-\infty}^{\infty} x_{\mathrm{p}}(t) x_{\mathrm{a}}(t)=0.\notag
\end{align}
$\bm{\mathrm{Proof.}}$
\begin{align}
	\label{eq:THM:orth:1}
	&\sum_{t=-\infty}^{\infty} x_{\mathrm{p}}(t) x_{\mathrm{a}}(t)
	=\sum_{\tau =0}^{\Pi -1} \sum_{k=-\infty}^{\infty} x_{\tau\mathrm{p}}(k) x_{\tau\mathrm{a}}(k),
\end{align}
where
\begin{align}
	\label{eq:THM:orth:sum}
	\sum_{k=-\infty}^{\infty} x_{\tau\mathrm{p}}(k) x_{\tau\mathrm{a}}(k)=
	\frac{1}{2}\sum_{k=-\infty}^{\infty}(x_{\tau\mathrm{p}}(k)+x_{\tau\mathrm{a}}(k))^2\notag\\
	-\frac{1}{2}\sum_{k=-\infty}^{\infty}x_{\tau\mathrm{p}}^2(k)
	-\frac{1}{2}\sum_{k=-\infty}^{\infty}x_{\tau\mathrm{a}}^2(k).
\end{align}
The Parseval's theorem
\begin{align}
	\sum_{k=-\infty}^{\infty} |x_{\tau}(k)|^2=\frac{1}{2\pi}\int_{-\pi}^{\pi}|X_{\tau}(\omega)|^2d\omega\notag
\end{align}
rewrites the terms in the discrete-time domain into
\begin{subequations}
	\label{eq:orth:234}
\begin{align}
	\label{eq:orth:2}
	&\sum_{k=-\infty}^{\infty} (x_{\tau\mathrm{p}}(k)+x_{\tau\mathrm{a}}(k))^2\notag\\
	&\hspace{2em}=\frac{1}{2\pi}\int_{-\pi}^{\pi}(X_{\tau\mathrm{p}}(\omega)+X_{\tau\mathrm{a}}(\omega))^2d\omega,\\
	\label{eq:orth:3}
	&\sum_{k=-\infty}^{\infty} x_{\tau\mathrm{p}}^2(k)=\frac{1}{2\pi}\int_{-\pi}^{\pi}X_{\tau\mathrm{p}}^2(\omega)d\omega,\\
	\label{eq:orth:4}
	&\sum_{k=-\infty}^{\infty} x_{\tau\mathrm{a}}^2(k)=\frac{1}{2\pi}\int_{-\pi}^{\pi}X_{\tau\mathrm{a}}^2(\omega)d\omega,
\end{align}
\end{subequations}
in the frequency domain.
According to \eqref{eq:DEF:x_tauap}, the lifted quasi-periodic-state and quasi-aperiodic-state functions satisfy $x_{\tau\mathrm{p}} \in \mathbb{S_A}^{\mathrm{c}}$ and $x_{\tau\mathrm{a}} \in \mathbb{S_P}^{\mathrm{c}}$, respectively.
Because of Lemma~\ref{LEM:FourierP/A}, $x_{\tau\mathrm{p}} \in \mathbb{S_A}^{\mathrm{c}}$, and $x_{\tau\mathrm{a}} \in \mathbb{S_P}^{\mathrm{c}}$, the Fourier transformed quasi-periodic and quasi-aperiodic states satisfy
\begin{align}
	&X_{\tau \mathrm{p}}(\omega) =
	\left\{
	\begin{array}{cl}
		X_{\tau \mathrm{p}}(\omega)&\mathrm{if}\ |\omega|\leq\rho\\
		0&\mathrm{if}\ \rho<|\omega|
	\end{array}
	\right. ,\notag\\
	&X_{\tau \mathrm{a}}(\omega) =
	\left\{
	\begin{array}{cl}
		0&\mathrm{if}\ |\omega|\leq\rho\\
		X_{\tau \mathrm{a}}(\omega)&\mathrm{if}\ \rho<|\omega|
	\end{array}
	\right. .\notag
\end{align}
They calculate \eqref{eq:orth:234} as follows:
\begin{align}
	&\sum_{k=-\infty}^{\infty} (x_{\tau\mathrm{p}}(k)+x_{\tau\mathrm{a}}(k))^2
	=\frac{1}{2\pi}\int_{-\rho}^{\rho}X_{\tau\mathrm{p}}^2(\omega)d\omega\notag\\
	&\hspace{2em}+\frac{1}{2\pi}\int_{\rho}^{\pi}X_{\tau\mathrm{a}}^2(\omega)d\omega
	+\frac{1}{2\pi}\int_{-\pi}^{-\rho}X_{\tau\mathrm{a}}^2(\omega)d\omega,\notag\\
	&\sum_{k=-\infty}^{\infty} x_{\tau\mathrm{p}}^2(k)=\frac{1}{2\pi}\int_{-\rho}^{\rho}X_{\tau\mathrm{p}}^2(\omega)d\omega,\notag\\
	&\sum_{k=-\infty}^{\infty} x_{\tau\mathrm{p}}^2(k)
	=\frac{1}{2\pi}\int_{\rho}^{\pi}X_{\tau\mathrm{a}}^2(\omega)d\omega
	+
	\frac{1}{2\pi}\int_{-\pi}^{-\rho}X_{\tau\mathrm{a}}^2(\omega)d\omega.\notag
\end{align}
The terms and \eqref{eq:THM:orth:sum} result in the orthogonality of the lifted quasi-periodic-state function $x_{\tau\mathrm{p}}$ and lifted quasi-aperiodic-state function $x_{\tau\mathrm{a}}$
\begin{align}
	\sum_{k=-\infty}^{\infty} x_{\tau\mathrm{p}}(k) x_{\tau\mathrm{a}}(k)=0.\notag
\end{align}
This orthogonality and \eqref{eq:THM:orth:1} yield the orthogonality of the quasi-periodic-state function $x_{\mathrm{p}}$ and quasi-aperiodic-state function $x_{\mathrm{a}}$
\begin{align}
	\sum_{t=-\infty}^{\infty} x_{\mathrm{p}}(t) x_{\mathrm{a}}(t)=0.\notag
\end{align}
\mbox{ }\vspace{-10mm}
\begin{flushright} $\blacksquare$ \end{flushright}

\begin{figure*}[t]
	\begin{center}
		\includegraphics[width=0.95\hsize]{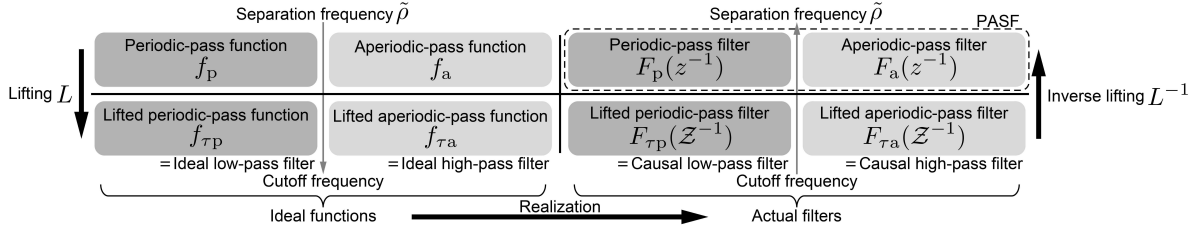}
	\caption{Realization flow of the PASF from the periodic-pass and aperiodic-pass functions to the periodic-pass and aperiodic-pass filters.}\label{fig:realization}
	\end{center}
\end{figure*}

\begin{HMpro}\label{PRO:pa}\end{HMpro}\vspace{-3mm}
The output of the state $x(t)$ by the periodic-pass function $f_{\mathrm{p}}$ and aperiodic-pass function $f_{\mathrm{a}}$ is zero
\begin{align}
	f_{\mathrm{a}}(f_{\mathrm{p}}(x(t)))=f_{\mathrm{p}}(f_{\mathrm{a}}(x(t)))=0.\notag
\end{align}
$\bm{\mathrm{Proof.}}$
According to \eqref{eq:DEF:fp} and \eqref{eq:DEF:fa},
\begin{align}
	&f_{\mathrm{p}}(x(t))\in \mathbb{S_A}^\mathrm{c},\
	f_{\mathrm{a}}(x(t))\in \mathbb{S_P}^\mathrm{c},\notag
\end{align}
and
\begin{align}
	f_{\mathrm{a}}(f_{\mathrm{p}}(x(t)))=f_{\mathrm{p}}(f_{\mathrm{a}}(x(t)))=0.\notag
\end{align}
\mbox{ }\vspace{-12mm}
\begin{flushright} $\blacksquare$ \end{flushright}


\section{Realization of Periodic/Aperiodic Separation Filter} \label{sec:3}
\subsection{Realization Framework} \label{sec:3-1}
Based on the definitions of the quasi-periodic and quasi-aperiodic states, this paper constructs causal linear time-invariant periodic-pass and aperiodic-pass filters that are an approximate realization of the periodic-pass function $f_{\mathrm{p}}$ in \eqref{eq:DEF:fp} and aperiodic-pass function $f_{\mathrm{a}}$ in \eqref{eq:DEF:fa}.
The PASF, which comprises the periodic-pass and aperiodic-pass filters, provides separated quasi-periodic state $\tilde{x}_{\mathrm{p}}(t)$ and quasi-aperiodic state $\tilde{x}_{\mathrm{a}}(t)$ from the periodic/aperiodic state $x_{\mathrm{pa}}(t)$.
As shown in Fig.~\ref{fig:realization}, the PASF is realized by lifting the periodic-pass and aperiodic-pass functions, realizing the ideal low-pass and high-pass filters as the lifted periodic-pass and aperiodic-pass filters, and inverse lifting the lifted filters into the periodic-pass and aperiodic-pass filters.
For the realization, this paper introduces two Z-transforms with $z$ and $\mathcal{Z}$, which are related as Proposition~\ref{PRO:Zz}.
\begin{HMpro}\label{PRO:Zz}\end{HMpro}\vspace{-3mm}
The $z$-transform and inverse $z$-transform for the state $x(t)$:
\begin{align}
	&\mathfrak{z}[x(t)] \coloneqq X(z^{-1}) \coloneqq \sum_{t=-\infty}^{\infty}x(t)z^{-t},\notag\\
	&\mathfrak{z}^{-1}[X(z^{-1})] \coloneqq \frac{1}{2\pi j}\oint_CX(z^{-1})z^{t-1}dz=x(t),\notag
\end{align}
and the $\mathcal{Z}$-transform and inverse $\mathcal{Z}$-transform for the lifted state $x_{\tau}(k)$:
\begin{align}
	&\mathscr{Z}[x_{\tau}(k)] \coloneqq X_{\tau}(\mathcal{Z}^{-1}) \coloneqq \sum_{k=-\infty}^{\infty}x_{\tau}(k)\mathcal{Z}^{-k},\notag\\
	&\mathscr{Z}^{-1}[X_{\tau}(\mathcal{Z}^{-1})] \coloneqq \frac{1}{2\pi j}\oint_CX_{\tau}(\mathcal{Z}^{-1})\mathcal{Z}^{k-1}d\mathcal{Z}=x_{\tau}(k),\notag
\end{align}
are related as
\begin{align}
	L^{-1}(\mathscr{Z}^{-1}[\mathcal{Z}^{-1}X_{\tau}(\mathcal{Z}^{-1})])
	= \mathfrak{z}^{-1}[z^{-\Pi}X(z^{-1})]
	= x(t-\Pi).\notag
\end{align}
$\bm{\mathrm{Proof.}}$
The inverse $\mathcal{Z}$-transform of $\mathcal{Z}^{-1}X_{\tau}(\mathcal{Z}^{-1})$ is
\begin{align}
	\mathscr{Z}^{-1}[\mathcal{Z}^{-1}X_{\tau}(\mathcal{Z}^{-1})]=x_{\tau}(k-1),\notag
\end{align}
which is inverse lifted by \eqref{eq:invlifting} as
\begin{align}
	L^{-1}(x_{\tau}(k-1)) = x((k-1)\Pi+\tau) = x(t-\Pi).\notag
\end{align}
Additionally, the inverse $z$-transform of $z^{-\Pi}X(z^{-1})$ is
\begin{align}
	\mathfrak{z}^{-1}[z^{-\Pi}X(z^{-1})] = x(t-\Pi).\notag
\end{align}
\mbox{ }\vspace{-12mm}
\begin{flushright} $\blacksquare$ \end{flushright}

In this paper, the lifted periodic-pass and aperiodic-pass functions are realized by caudal linear filters as lifted periodic-pass and aperiodic-pass filters
\begin{align}
	&\tilde{x}_{\tau\mathrm{p}}(k) = - \sum_{i=1}^{N} a_i\tilde{x}_{\tau\mathrm{p}}(k-i) + \sum_{i=0}^{N} b_ix_{\tau \mathrm{pa}}(k-i),\notag\\
	&\tilde{x}_{\tau\mathrm{a}}(k) = - \sum_{i=1}^{N} c_i\tilde{x}_{\tau\mathrm{a}}(k-i) + \sum_{i=0}^{N} d_ix_{\tau \mathrm{pa}}(k-i),\notag\\
	&a_i,\ b_i,\ c_i,\ d_i\in \mathbb{R},\ N\in \mathbb{Z}_{>0},\notag
\end{align}
respectively.
Then, the filters are $\mathcal{Z}$-transformed with respect to $k$ into
\begin{align}
	\tilde{X}_{\tau\mathrm{p}}(\mathcal{Z}^{-1}) &=F_{\tau\mathrm{p}}(\mathcal{Z}^{-1}) X_{\tau \mathrm{pa}}(\mathcal{Z}^{-1}),\notag\\
	\tilde{X}_{\tau\mathrm{a}}(\mathcal{Z}^{-1}) &=F_{\tau\mathrm{a}}(\mathcal{Z}^{-1}) X_{\tau \mathrm{pa}}(\mathcal{Z}^{-1}),\notag\\
	F_{\tau\mathrm{p}}(\mathcal{Z}^{-1}) &\coloneqq \frac{b_0+b_1\mathcal{Z}^{-1}+ \ldots +b_{N}\mathcal{Z}^{-N}}{1+a_1\mathcal{Z}^{-1}+ \ldots +a_{N}\mathcal{Z}^{-N}},\notag\\
	F_{\tau\mathrm{a}}(\mathcal{Z}^{-1}) &\coloneqq \frac{d_0+d_1\mathcal{Z}^{-1}+ \ldots +d_{N}\mathcal{Z}^{-N}}{1+c_1\mathcal{Z}^{-1}+ \ldots +c_{N}\mathcal{Z}^{-N}}.\notag
\end{align}
The coefficients $a_i$ and $b_i$ are determined to construct $F_{\tau\mathrm{p}}(\mathcal{Z}^{-1})$ to be a low-pass filter; $c_i$ and $d_i$ are determined to construct $F_{\tau\mathrm{a}}(\mathcal{Z}^{-1})$ to be a high-pass filter.
According to Proposition~\ref{PRO:Zz}, the inverse $\mathcal{Z}$-transform and inverse lifting function $L^{-1}$ derive the periodic-pass and aperiodic-pass filters:
\begin{align}
	\tilde{x}_{\mathrm{p}}(t)=& - \sum_{i=1}^{N} a_i\tilde{x}_{\mathrm{p}}(t-i\Pi) + \sum_{i=0}^{N} b_ix_{\mathrm{pa}}(t-i\Pi),\notag\\
	\tilde{x}_{\mathrm{a}}(t)=& - \sum_{i=1}^{N} c_i\tilde{x}_{\mathrm{a}}(t-i\Pi) + \sum_{i=0}^{N} d_ix_{\mathrm{pa}}(t-i\Pi),\notag
\end{align}
which are the PASF.
Additionally, the $z$-transformed periodic-pass filter $F_{\mathrm{p}}(z^{-1})$ and aperiodic-pass filter $F_{\mathrm{a}}(z^{-1})$ are
\begin{align}
	\tilde{X}_{\mathrm{p}}(z^{-1}) &=F_{\mathrm{p}}(z^{-1}) X_{\mathrm{pa}}(z^{-1}),\notag\\
	\tilde{X}_{\mathrm{a}}(z^{-1}) &=F_{\mathrm{a}}(z^{-1}) X_{\mathrm{pa}}(z^{-1}),\notag\\
	F_{\mathrm{p}}(z^{-1}) &= \frac{b_0+b_1z^{-\Pi}+ \ldots +b_{N}z^{-N\Pi}}{1+a_1z^{-\Pi}+ \ldots +a_{N}z^{-N\Pi}},\notag\\
	F_{\mathrm{a}}(z^{-1}) &= \frac{d_0+d_1z^{-\Pi}+ \ldots +d_{N}z^{-N\Pi}}{1+c_1z^{-\Pi}+ \ldots +c_{N}z^{-N\Pi}}.\notag
\end{align}
The limitation of the realization is that the filters cannot ideally realize the periodic-pass and aperiodic-pass functions owing to the impossibility of realizing ideal and causal low-pass and high-pass filters.
This realization error results in the interference of the periodic and aperiodic states between the separated quasi-periodic and quasi-aperiodic states.

\begin{figure}[t]
	\begin{center}
		\includegraphics[width=\hsize]{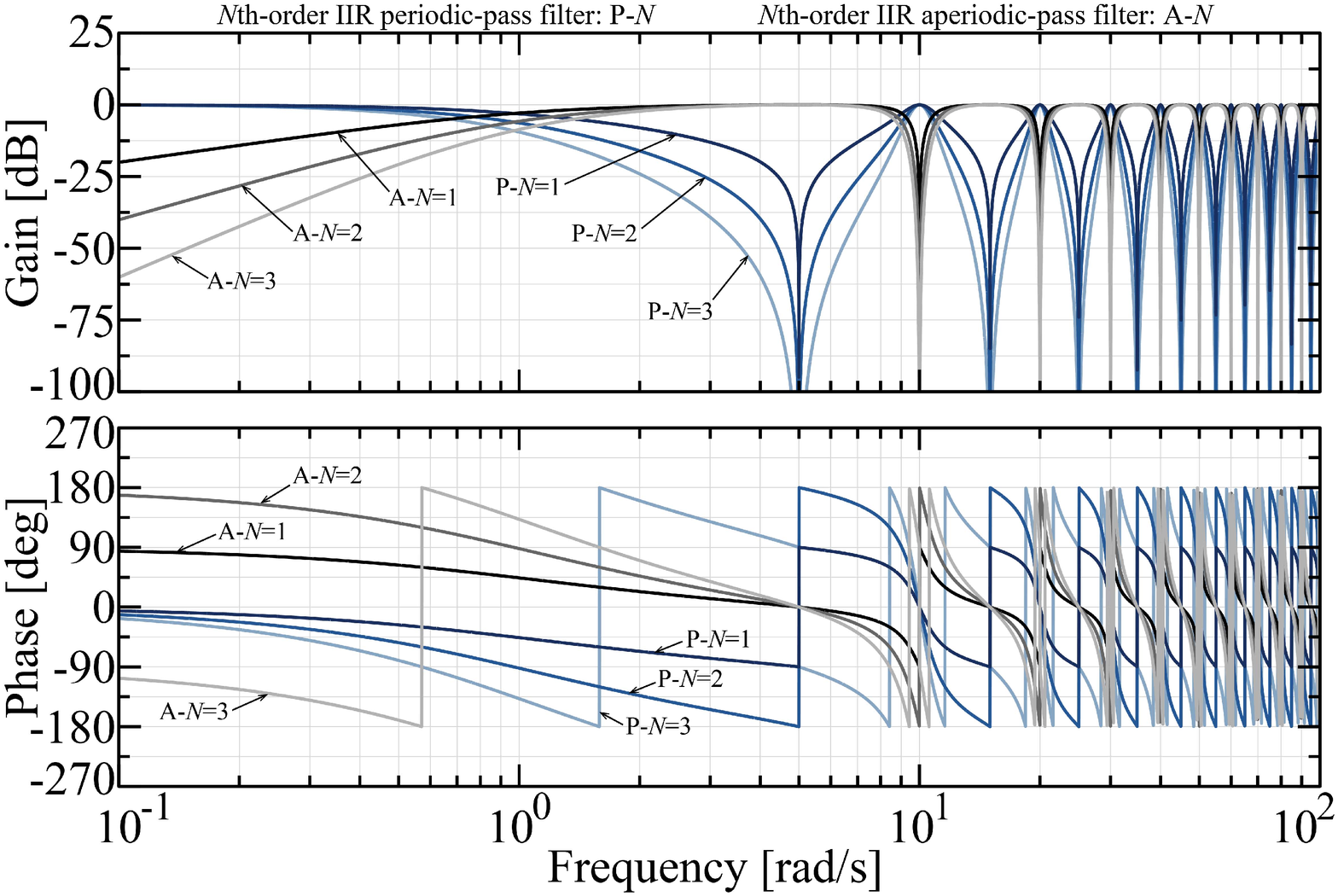}
		\caption{Bode plots of the $N$th-order IIR periodic-pass filter $F_{\mathrm{p}}(z^{-1})$ and aperiodic-pass filter $F_{\mathrm{p}}(z^{-1})$, which use the sampling time $T=0.001\ \mathrm{s}$, period $\Pi=628$, and separation frequency $\tilde{\rho}=1\ \mathrm{rad/s}$.}\label{fig:NthIIR}
	\end{center}
\end{figure}

\begin{figure}[t]
	\begin{center}
		\includegraphics[width=\hsize]{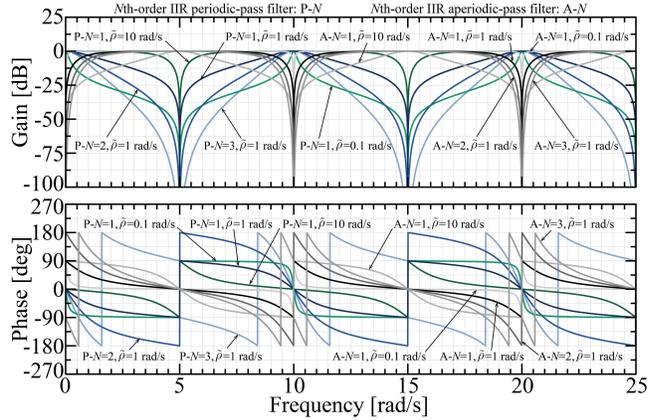}
	\caption{Different effects of the order $N$ and separation frequency $\tilde{\rho}$ on the frequency characteristics of the periodic-pass filter $F_{\mathrm{p}}(z^{-1})$ and aperiodic-pass filter $F_{\mathrm{p}}(z^{-1})$, which use the sampling time $T=0.001\ \mathrm{s}$ and period $\Pi=628$.}\label{fig:NthIIRrho}
	\end{center}
\end{figure}

\subsection{IIR Realization} \label{sec:3-2}
Consider IIR periodic-pass and aperiodic-pass filters.
The lifted periodic-pass and aperiodic-pass filters can utilize $N$th-order IIR low-pass and high-pass filters as
\begin{align}
	F_{\tau\mathrm{p}}(\mathcal{Z}^{-1})=\left(\frac{\tilde{\rho}}{\tilde{s}+\tilde{\rho}}\right)^N,\
	F_{\tau\mathrm{a}}(\mathcal{Z}^{-1})=\left(\frac{\tilde{s}}{\tilde{s}+\tilde{\rho}}\right)^N,\notag
\end{align}
where $\tilde{s}$ is the $\mathcal{Z}$-transformed approximate representation of the Laplace operator by the bilinear transform with $\mathcal{Z}$:
\begin{align}
	\tilde{s} \coloneqq \frac{2}{\Pi T} \frac{1-\mathcal{Z}^{-1}}{1+\mathcal{Z}^{-1}}.\notag
\end{align}
The separation frequency $\tilde{\rho}$ is used as the cutoff frequency according to the definition in \eqref{eq:DEF:sets} with \eqref{eq:rho}.
Using Proposition~\ref{PRO:Zz}, the IIR periodic-pass and aperiodic-pass filters are derived as
\begin{subequations}
	\label{eq:IIR:LowHigh:z1}
\begin{align}
	\label{eq:IIR:LowHigh:z1:P}
	F_{\mathrm{p}}(z^{-1})&=\left(\frac{\tilde{\rho}\Pi T(1+z^{-\Pi})}{2(1-z^{-\Pi})+\tilde{\rho}\Pi T(1+z^{-\Pi})}\right)^N,\\
	\label{eq:IIR:LowHigh:z1:A}
	F_{\mathrm{a}}(z^{-1})&=\left(\frac{2(1-z^{-\Pi})}{2(1-z^{-\Pi})+\tilde{\rho}\Pi T(1+z^{-\Pi})}\right)^N.
\end{align}
\end{subequations}
Fig.~\ref{fig:NthIIR} shows the Bode plots of the $N$th-order IIR periodic-pass filter $F_{\mathrm{p}}(z^{-1})$ and aperiodic-pass filter $F_{\mathrm{a}}(z^{-1})$, where the increase in the order deepens the band-stop characteristics.
Moreover, Fig.~\ref{fig:NthIIRrho} shows the filters with variations in the separation frequency $\tilde{\rho}$ in addition to the variations in the order $N$, where the increase in the separation frequency $\tilde{\rho}$ extends the band-pass bandwidth of the periodic-pass filter $F_{\mathrm{p}}(z^{-1})$.
The order and calculation cost of these IIR filters can be lower than those of the FIR filters.

\begin{figure}[t]
	\begin{center}
		\includegraphics[width=\hsize]{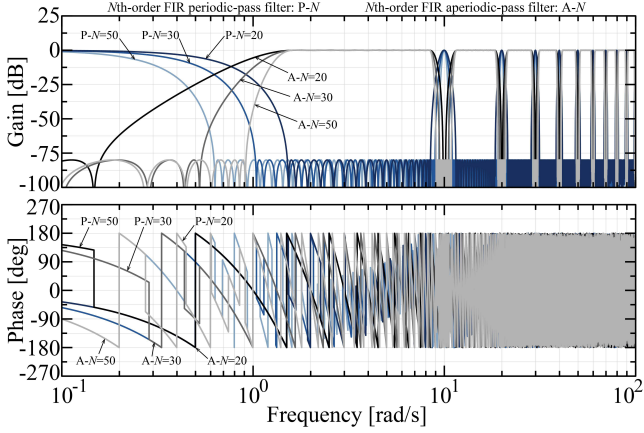}
		\caption{Bode plots of the $N$th-order FIR periodic-pass filter $F_{\mathrm{p}}(z^{-1})$ and aperiodic-pass filter $F_{\mathrm{a}}(z^{-1})$, which use the sampling time $T=0.001\ \mathrm{s}$, period $\Pi=628$, and separation frequency $\tilde{\rho}=1\ \mathrm{rad/s}$.}\label{fig:NthFIR}
	\end{center}
\end{figure}

\subsection{FIR Realization} \label{sec:3-3}
Consider FIR periodic-pass and aperiodic-pass filters.
In contrast to the IIR filters, the FIR filters set $a_i$ and $c_i$ of the denominator polynomials to zero; hence, the FIR filters are inherently stable.
This study designed three equi-ripple FIR low-pass and high-pass filters using the Parks-McClellan algorithm for the lifted periodic-pass filter $F_{\tau\mathrm{p}}(\mathcal{Z}^{-1})$ and the lifted aperiodic-pass filter $F_{\tau\mathrm{a}}(\mathcal{Z}^{-1})$, where MATLAB function \textit{firceqrip()} was used to calculate the coefficients $b_i$ and $d_i$ of the 20th, 30th, and 50th FIR low-pass and high-pass filters.
Fig.~\ref{fig:NthFIR} shows the Bode plots of the $N$th-order FIR periodic-pass filter $F_{\mathrm{p}}(z^{-1})$ and aperiodic-pass filter $F_{\mathrm{a}}(z^{-1})$, where the slope increases as the order increases.
The filter gain has a steeper slope with a higher order than those of the IIR filters in Fig.~\ref{fig:NthIIR}.
Nevertheless, the phase of the FIR aperiodic-pass filters has a lag at the band-pass frequencies; hence, an aperiodic state output by the FIR aperiodic-pass filter lags.
This problem can be solved by a complementary realization of the aperiodic-pass function.

\subsection{Complementary Realization} \label{sec:3-4}
\begin{figure}[t]
	\begin{center}
		\includegraphics[width=\hsize]{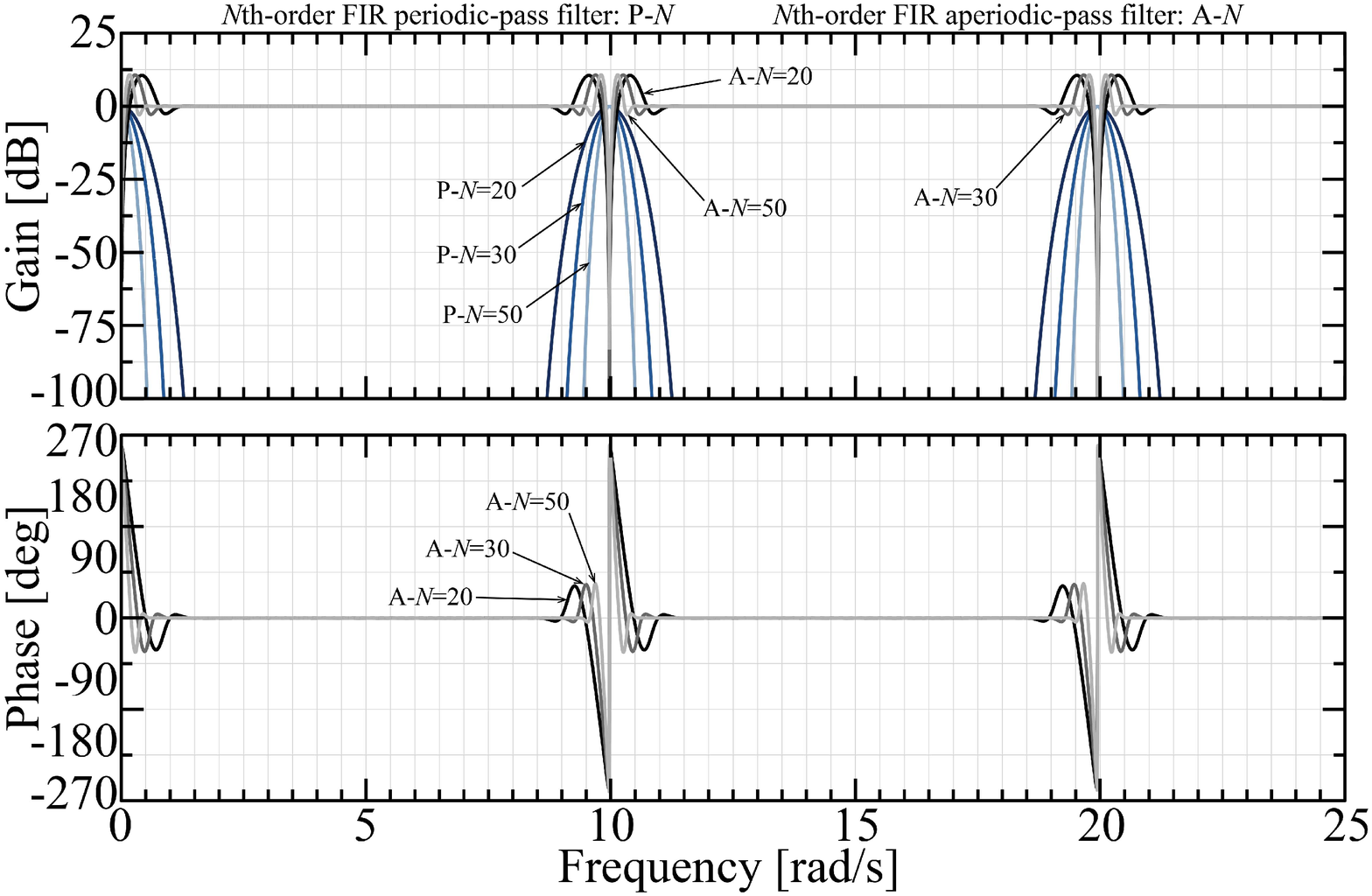}
		\caption{Complementary realization of the FIR aperiodic-pass filter $F_{\mathrm{a}}(z^{-1})$ that improves the phase-lag of the aperiodic-pass filter at the band-pass frequencies. It uses the sampling time $T=0.001\ \mathrm{s}$, period $\Pi=628$, and separation frequency $\tilde{\rho}=1\ \mathrm{rad/s}$.}\label{fig:NthFIR:comp}
	\end{center}
\end{figure}

The complementary realization designs the aperiodic-pass filter to be a complementary filter of the periodic-pass filter as
\begin{align}
	F_{\mathrm{a}}(z^{-1}) = 1-F_{\mathrm{p}}(z^{-1}).\notag
\end{align}
The first-order IIR periodic-pass and aperiodic-pass filters based on \eqref{eq:IIR:LowHigh:z1:P} and \eqref{eq:IIR:LowHigh:z1:A} are the complementary filters as
\begin{align}
	F_{\mathrm{p}}(z^{-1})&=\frac{\tilde{\rho}\Pi T(1+z^{-\Pi})}{2(1-z^{-\Pi})+\tilde{\rho}\Pi T(1+z^{-\Pi})},\notag\\
	F_{\mathrm{a}}(z^{-1})&=\frac{2(1-z^{-\Pi})}{2(1-z^{-\Pi})+\tilde{\rho}\Pi T(1+z^{-\Pi})}.\notag
\end{align}
The complementary realization can improve the phase lag of the FIR aperiodic-pass function.
Compared to the phase of the FIR aperiodic-pass filters in Fig.~\ref{fig:NthFIR}, that of the complementary FIR aperiodic-pass filters is zero at the band-pass frequencies, as shown in Fig.~\ref{fig:NthFIR:comp}.

\section{Periodic/Aperiodic Separation Filter with Kalman Filter} \label{sec:4}
\subsection{Preliminaries} \label{sec:4-1}
Consider an observable linear time-invariant system:
\begin{subequations}
	\label{eq:DEF:LTIsys}
\begin{align}
	&\bm{x}_{\mathrm{pa}}(t+1)=\bm{A}\bm{x}_{\mathrm{pa}}(t)+\bm{B}\bm{u}(t)+\bm{v}(t),\\
	&\bm{y}(t)=\bm{C}\bm{x}_{\mathrm{pa}}(t)+\bm{w}(t),\\
	&\mathrm{E}[\bm{v}(t)] = \bm{0},\
	\mathrm{E}[\bm{w}(t)] = \bm{0},\\
	&\bm{Q}\coloneqq\mathrm{E}[\bm{v}(t)\bm{v}^{\mathrm{T}}(t)],\
	\bm{R}\coloneqq\mathrm{E}[\bm{w}(t)\bm{w}^{\mathrm{T}}(t)],\notag\\
	&\bm{x}_{\mathrm{pa}}(t), \bm{v}(t)\in\mathbb{R}^{n},\
	\bm{u}(t)\in\mathbb{R}^{p},\
	\bm{y}(t), \bm{w}(t)\in\mathbb{R}^{m},\notag\\
	&\bm{A}, \bm{Q}, \bm{R}\in\mathbb{R}^{n \times n},\
	\bm{B}\in\mathbb{R}^{n \times p},\ \bm{C}\in\mathbb{R}^{m \times n},\notag
\end{align}
\end{subequations}
the Kalman filter:
\begin{subequations}
	\label{eq:PRE:KF}
\begin{align}
	&\hat{\bm{x}}_{\mathrm{pa}}(t|t-1) = \bm{A}\hat{\bm{x}}_{\mathrm{pa}}(t-1|t-1) + \bm{B}\bm{u}(t-1),\\
	&\bm{P}(t|t-1)=\bm{A}\bm{P}(t-1|t-1)\bm{A}^{\mathrm{T}}+\bm{Q},\\
	&\bm{g}(t) = \bm{P}(t|t-1)\bm{C}^{\mathrm{T}}[\bm{C}\bm{P}(t|t-1)\bm{C}^{\mathrm{T}} + \bm{R}]^{-1},\\
	&\hat{\bm{x}}_{\mathrm{pa}}(t|t) = \hat{\bm{x}}_{\mathrm{pa}}(t|t-1) + \bm{g}(t)[\bm{y}(t) - \bm{C}\hat{\bm{x}}_{\mathrm{pa}}(t|t-1)],\\
	&\bm{P}(t|t)=(\bm{I} - \bm{g}(t)\bm{C})\bm{P}(t|t-1),\\
	&\hat{\bm{x}}_{\mathrm{pa}}(t)\in\mathbb{R}^{n},\ \bm{P}(t|t)\in\mathbb{R}^{n},\ \bm{g}(t)\in\mathbb{R}^{n \times m},\notag
\end{align}
\end{subequations}
and the PASF:
\begin{subequations}
	\label{eq:PRE:PASF}
\begin{align}
	&\tilde{\bm{x}}_{\mathrm{p}}(t)= \tilde{\bm{\theta}}_{\mathrm{p}}(t-\Pi) + \bm{S}_{\mathrm{p}}\bm{x}_{\mathrm{pa}}(t),\\
	&\tilde{\bm{x}}_{\mathrm{a}}(t)= \tilde{\bm{\theta}}_{\mathrm{a}}(t-\Pi) + \bm{S}_{\mathrm{a}}\bm{x}_{\mathrm{pa}}(t),\\
	&\tilde{\bm{\theta}}_{\mathrm{p}}(t-\Pi)\coloneqq\displaystyle\sum_{i=1}^{N}[\bm{G}_{\mathrm{p}i}\tilde{\bm{x}}_{\mathrm{p}}(t-i\Pi) + \bm{H}_{\mathrm{p}i}\bm{x}_{\mathrm{pa}}(t-i\Pi)],\\
	&\tilde{\bm{\theta}}_{\mathrm{a}}(t-\Pi)\coloneqq\displaystyle\sum_{i=1}^{N}[\bm{G}_{\mathrm{a}i}\tilde{\bm{x}}_{\mathrm{a}}(t-i\Pi) + \bm{H}_{\mathrm{a}i}\bm{x}_{\mathrm{pa}}(t-i\Pi)],\\
	&\bm{G}_{\mathrm{p}i}\coloneqq\mathrm{diag}(-a_{i1},\ \cdots,\ -a_{in}),\notag\\
	&\bm{G}_{\mathrm{a}i}\coloneqq\mathrm{diag}(-c_{i1},\ \cdots,\ -c_{in}),\notag\\
	&\bm{H}_{\mathrm{p}i}\coloneqq\mathrm{diag}(b_{i1},\ \cdots,\ a_{in}),\
	\bm{H}_{\mathrm{a}i}\coloneqq\mathrm{diag}(d_{i1},\ \cdots,\ d_{in}),\notag\\
	&\bm{S}_{\mathrm{p}}\coloneqq\mathrm{diag}(b_{01},\ \cdots,\ a_{0n}),\
	\bm{S}_{\mathrm{a}}\coloneqq\mathrm{diag}(d_{01},\ \cdots,\ d_{0n}),\notag\\
	&\tilde{\bm{x}}_{\mathrm{p}}(t), \tilde{\bm{x}}_{\mathrm{a}}(t), \tilde{\bm{\theta}}_{\mathrm{p}}(t), \tilde{\bm{\theta}}_{\mathrm{a}}(t)\in\mathbb{R}^{n}\notag\\
	&\bm{S}_{\mathrm{p}}, \bm{S}_{\mathrm{a}}, \bm{G}_{\mathrm{p}i}, \bm{G}_{\mathrm{a}i}, \bm{H}_{\mathrm{p}i}, \bm{H}_{\mathrm{a}i}\in\mathbb{R}^{n \times n}.\notag
\end{align}
\end{subequations}
Note that the state of the system is supposed to be the periodic/aperiodic state.
The vectors $\hat{\bm{x}}_{\mathrm{pa}}(t|t-1)$, $\hat{\bm{x}}_{\mathrm{pa}}(t|t)$, $\bm{g}(t)$, $\bm{v}(t)$, and $\bm{w}(t)$ denote the predicted periodic/aperiodic state, updated periodic/aperiodic state, Kalman gain, process noise, and observation noise, respectively.
The matrices $\bm{G}_{\mathrm{p}i}$, $\bm{G}_{\mathrm{a}i}$, $\bm{H}_{\mathrm{p}i}$, $\bm{H}_{\mathrm{a}i}$, $\bm{S}_{\mathrm{p}i}$, and $\bm{S}_{\mathrm{a}i}$ are composed of the coefficients of the periodic-pass and aperiodic-pass filters.
The estimation errors are defined by
\begin{align}
	\label{eq:DEF:KFerror}
	&\bm{e}(t|t) \coloneqq \bm{x}_{\mathrm{pa}}(t) - \hat{\bm{x}}_{\mathrm{pa}}(t|t),\\
	&\bm{e}_{\mathrm{p}}(t|t) \coloneqq \bm{x}_{\mathrm{p}}(t) - \hat{\bm{x}}_{\mathrm{p}}(t|t),\
	\bm{e}_{\mathrm{a}}(t|t) \coloneqq \bm{x}_{\mathrm{a}}(t) - \hat{\bm{x}}_{\mathrm{a}}(t|t),\notag\\
	&\bm{e}(t|t), \bm{e}_{\mathrm{p}}(t|t), \bm{e}_{\mathrm{a}}(t|t)\in\mathbb{R}^{n},\notag
\end{align}
with $\bm{e}(t|t)$, $\bm{e}_{\mathrm{p}}(t|t)$, and $\bm{e}_{\mathrm{a}}(t|t)$ denoting the estimation error, periodic-estimation error, and aperiodic-estimation error, respectively.
The error covariance matrix is defined by
\begin{align}
	\bm{P}(t|t) &\coloneqq \mathrm{E}[\bm{e}(t|t)\bm{e}^{\mathrm{T}}(t|t)].\notag
\end{align}
To demonstrate that the KF-PASF achieves unbiased estimation and sum minimization of periodic-error and aperiodic-error covariances, the following is assumed.
\begin{HMassum}\label{ASS:ideal}\end{HMassum}\vspace{-3mm}
The PASF in \eqref{eq:PRE:PASF} satisfies \eqref{eq:DEF:fp} and \eqref{eq:DEF:fa} as
\begin{align}
	\tilde{\bm{x}}_{\mathrm{p}}(t)=\bm{x}_{\mathrm{p}}(t),\
	\tilde{\bm{x}}_{\mathrm{a}}(t)=\bm{x}_{\mathrm{a}}(t),\
	\bm{x}_{\mathrm{p}}(t), \bm{x}_{\mathrm{a}}(t)\in\mathbb{R}^{n},\notag
\end{align}
which gives
\begin{align}
	\tilde{\bm{\theta}}_{\mathrm{p}}(t)=\bm{\theta}_{\mathrm{p}}(t),\
	\tilde{\bm{\theta}}_{\mathrm{a}}(t)=\bm{\theta}_{\mathrm{a}}(t),\
	{\bm{\theta}}_{\mathrm{p}}(t), {\bm{\theta}}_{\mathrm{a}}(t)\in\mathbb{R}^{n },\notag
\end{align}
where
\begin{align*}
	{\bm{\theta}}_{\mathrm{p}}(t-\Pi)\coloneqq\displaystyle\sum_{i=1}^{N}[\bm{G}_{\mathrm{p}i}{\bm{x}}_{\mathrm{p}}(t-i\Pi) + \bm{H}_{\mathrm{p}i}\bm{x}_{\mathrm{pa}}(t-i\Pi)],\\
	{\bm{\theta}}_{\mathrm{a}}(t-\Pi)\coloneqq\displaystyle\sum_{i=1}^{N}[\bm{G}_{\mathrm{a}i}{\bm{x}}_{\mathrm{a}}(t-i\Pi) + \bm{H}_{\mathrm{a}i}\bm{x}_{\mathrm{pa}}(t-i\Pi)].
\end{align*}

If Assumption~\ref{ASS:ideal} holds, the expectation of the product of the periodic-estimation and aperiodic-estimation errors is zero based on Theorem~\ref{THM:orthogonality}
\begin{align}
	\label{eq:assume:ortho_PA_epea}
	\mathrm{E}[\bm{e}_{\mathrm{p}}(t|t)\bm{e}_{\mathrm{a}}^{\mathrm{T}}(t|t)]&=\sum_{t=-\infty}^{\infty}\bm{e}_{\mathrm{p}}(t|t)\bm{e}_{\mathrm{a}}^{\mathrm{T}}(t|t)=\bm{0}.
\end{align}

\subsection{Algorithm, Unbiased Estimation, and Equivalent Sum of Covariances} \label{sec:4-2}
The Kalman filter in \eqref{eq:PRE:KF} and the PASF in \eqref{eq:PRE:PASF} are integrated into the KF-PASF in Algorithm~\ref{alg}, where the predicted quasi-periodic state $\hat{\bm{x}}_{\mathrm{p}}(t|t-1)$, predicted quasi-aperiodic state $\hat{\bm{x}}_{\mathrm{a}}(t|t-1)$, updated quasi-periodic state $\hat{\bm{x}}_{\mathrm{p}}(t|t)$, and updated quasi-aperiodic state $\hat{\bm{x}}_{\mathrm{a}}(t|t)$ are obtained as
\begin{subequations}
	\label{eq:DEF:estimator:update:PA}
\begin{align}
	&\hat{\bm{x}}_{\mathrm{p}}(t|t-1)= \hat{\bm{\theta}}_{\mathrm{p}}(t-\Pi|t-\Pi) + \bm{S}_{\mathrm{p}}\hat{\bm{x}}_{\mathrm{pa}}(t|t-1),\notag\\
	&\hat{\bm{x}}_{\mathrm{a}}(t|t-1)= \hat{\bm{\theta}}_{\mathrm{a}}(t-\Pi|t-\Pi) + \bm{S}_{\mathrm{a}}\hat{\bm{x}}_{\mathrm{pa}}(t|t-1),\notag\\
	&\hat{\bm{x}}_{\mathrm{p}}(t|t)= \hat{\bm{\theta}}_{\mathrm{p}}(t-\Pi|t-\Pi) + \bm{S}_{\mathrm{p}}\hat{\bm{x}}_{\mathrm{pa}}(t|t),\\
	&\hat{\bm{x}}_{\mathrm{a}}(t|t)= \hat{\bm{\theta}}_{\mathrm{a}}(t-\Pi|t-\Pi) + \bm{S}_{\mathrm{a}}\hat{\bm{x}}_{\mathrm{pa}}(t|t),\\
	&\hat{\bm{x}}_{\mathrm{p}}(t|t), \hat{\bm{x}}_{\mathrm{a}}(t|t), \hat{\bm{\theta}}_{\mathrm{p}}(t), \hat{\bm{\theta}}_{\mathrm{a}}(t)\in\mathbb{R}^{n}.\notag
\end{align}
\end{subequations}
where
\begin{align*}
	&\hat{\bm{\theta}}_{\mathrm{p}}(t-\Pi|t-\Pi)\coloneqq\\
	&\displaystyle\sum_{i=1}^{N}[\bm{G}_{\mathrm{p}i}\hat{\bm{x}}_{\mathrm{p}}(t-i\Pi|t-i\Pi) + \bm{H}_{\mathrm{p}i}\hat{\bm{x}}_{\mathrm{pa}}(t-i\Pi|t-i\Pi)]\\
	&\hat{\bm{\theta}}_{\mathrm{a}}(t-\Pi|t-\Pi)\coloneqq\\
	&\displaystyle\sum_{i=1}^{N}[\bm{G}_{\mathrm{a}i}\hat{\bm{x}}_{\mathrm{a}}(t-i\Pi|t-i\Pi) + \bm{H}_{\mathrm{a}i}\hat{\bm{x}}_{\mathrm{pa}}(t-i\Pi|t-i\Pi)].
\end{align*}
The estimation errors of $\hat{\bm{\theta}}_{\mathrm{p}}(t-\Pi|t-\Pi)$ and $\hat{\bm{\theta}}_{\mathrm{a}}(t-\Pi|t-\Pi)$ are respectively defined as
\begin{align}
	&\bm{\epsilon}_{\mathrm{p}}(t-\Pi|t-\Pi) \coloneqq {\bm{\theta}}_{\mathrm{p}}(t-\Pi) - \hat{\bm{\theta}}_{\mathrm{p}}(t-\Pi|t-\Pi),\notag\\
	&\bm{\epsilon}_{\mathrm{a}}(t-\Pi|t-\Pi) \coloneqq {\bm{\theta}}_{\mathrm{a}}(t-\Pi) - \hat{\bm{\theta}}_{\mathrm{a}}(t-\Pi|t-\Pi),\notag\\
	&\bm{\epsilon}_{\mathrm{p}}(t|t), \bm{\epsilon}_{\mathrm{a}}(t|t), \in\mathbb{R}^{n}.\notag
\end{align}
For the KF-PASF, Theorems~\ref{THM:ub} and \ref{THM:ecm} show the unbiased estimation and the equivalent sum of the periodic-error and aperiodic-error covariances, respectively.
\begin{algorithm}[t]
\caption{KF-PASF}
\label{alg}
{
{\textbf Parameters and initialization:}
\vspace{-1em}
\begin{align*}
	&\bm{P}(0|0), \bm{Q},\ \bm{R},\ \Pi\\
	&\hat{\bm{x}}_{\mathrm{pa}}(-n|-n)=\mathrm{E}[\bm{x}_{\mathrm{pa}}(-n)],\
	\hat{\bm{x}}_{\mathrm{p}}(-n|-n)=\mathrm{E}[\bm{x}_{\mathrm{p}}(-n)]\\
	&\hat{\bm{x}}_{\mathrm{a}}(-n|-n)=\mathrm{E}[\bm{x}_{\mathrm{a}}(-n)],\
	n =0,\ 1,\ \ldots ,\ N\Pi-1
\end{align*}
\vspace{-2em}\\
{\textbf Prediction:}
\vspace{-1em}
\begin{align*}
	&\hat{\bm{x}}_{\mathrm{pa}}(t|t-1) = \bm{A}\hat{\bm{x}}_{\mathrm{pa}}(t-1|t-1) + \bm{B}\bm{u}(t-1)\\
	&\bm{\theta}_{\mathrm{p}}(t-\Pi|t-\Pi)=\\
	&\sum_{i=1}^{N}[\bm{G}_{\mathrm{p}i}\hat{\bm{x}}_{\mathrm{p}}(t-i\Pi|t-i\Pi) + \bm{H}_{\mathrm{p}i}\hat{\bm{x}}_{\mathrm{pa}}(t-i\Pi|t-i\Pi)]\\
	&\bm{\theta}_{\mathrm{a}}(t-\Pi|t-\Pi)=\\
	&\sum_{i=1}^{i}[\bm{G}_{\mathrm{a}i}\hat{\bm{x}}_{\mathrm{a}}(t-i\Pi|t-i\Pi) + \bm{H}_{\mathrm{a}i}\hat{\bm{x}}_{\mathrm{pa}}(t-i\Pi|t-i\Pi)]\\
	&\hat{\bm{x}}_{\mathrm{p}}(t|t-1)= \bm{\theta}_{\mathrm{p}}(t-\Pi|t-\Pi) + \bm{S}_{\mathrm{p}}\hat{\bm{x}}_{\mathrm{pa}}(t|t-1)\\
	&\hat{\bm{x}}_{\mathrm{a}}(t|t-1)= \bm{\theta}_{\mathrm{a}}(t-\Pi|t-\Pi) + \bm{S}_{\mathrm{a}}\hat{\bm{x}}_{\mathrm{pa}}(t|t-1)\\
	&\bm{P}(t|t-1)=\bm{A}\bm{P}(t-1|t-1)\bm{A}^{\mathrm{T}}+\bm{Q}
\end{align*}
\vspace{-2em}\\
{\textbf Update:}
\vspace{-1em}
\begin{align*}
	&\bm{g}(t) = \bm{P}(t|t-1)\bm{C}^{\mathrm{T}}[\bm{C}\bm{P}(t|t-1)\bm{C}^{\mathrm{T}} + \bm{R}]^{-1}\\
	&\hat{\bm{x}}_{\mathrm{pa}}(t|t) = \hat{\bm{x}}_{\mathrm{pa}}(t|t-1) + \bm{g}(t)[\bm{y}(t) - \bm{C}\hat{\bm{x}}_{\mathrm{pa}}(t|t-1)]\\
	&\hat{\bm{x}}_{\mathrm{p}}(t|t)= \bm{\theta}_{\mathrm{p}}(t-\Pi|t-\Pi) + \bm{S}_{\mathrm{p}}\hat{\bm{x}}_{\mathrm{pa}}(t|t) \\
	&\hat{\bm{x}}_{\mathrm{a}}(t|t)= \bm{\theta}_{\mathrm{a}}(t-\Pi|t-\Pi) + \bm{S}_{\mathrm{a}}\hat{\bm{x}}_{\mathrm{pa}}(t|t)\\
	&\bm{P}(t|t)=(\bm{I} - \bm{g}(t)\bm{C})\bm{P}(t|t-1)
\end{align*}
\vspace{-2em}
}
\end{algorithm}
\begin{HMthm}\label{THM:ub}\end{HMthm}\vspace{-3mm}
Assume Assumption~\ref{ASS:ideal} holds.
The estimation error $\bm{e}(t|t)$, periodic-estimation error $\bm{e}_{\mathrm{p}}(t|t)$, and aperiodic-estimation error $\bm{e}_{\mathrm{a}}(t|t)$ of the KF-PASF in Algorithm~\ref{alg} are unbiased if the initial errors are unbiased
\begin{align}
	&\mathrm{E}[\bm{e}(-n|-n)]=\mathrm{E}[\bm{e}_{\mathrm{p}}(-n|-n)]=\mathrm{E}[\bm{e}_{\mathrm{a}}(-n|-n)]=\bm{0},\notag\\
	&n =0,\ 1,\ \ldots ,\ N\Pi-1,\notag\\
	&\hspace{1em}\Rightarrow\ \forall t\in \mathbb{Z}_{>0},\ \mathrm{E}[\bm{e}(t|t)]=\mathrm{E}[\bm{e}_{\mathrm{p}}(t|t)]=\mathrm{E}[\bm{e}_{\mathrm{a}}(t|t)]=\bm{0}.\notag
\end{align}
$\bm{\mathrm{Proof.}}$
The dynamics of the expectation of the estimation error are
\begin{align}
	\label{eq:ub:ee}
	\mathrm{E}[\bm{e}(t|t)] = (\bm{I} - \bm{g}(t)\bm{C})\bm{A}\mathrm{E}[\bm{e}(t-1|t-1)],
\end{align}
which can be derived from \eqref{eq:DEF:LTIsys}, \eqref{eq:PRE:KF}, and \eqref{eq:DEF:KFerror}.
The PASF for the periodic/aperiodic state in \eqref{eq:PRE:PASF}, the PASF for the estimated periodic/aperiodic state in \eqref{eq:DEF:estimator:update:PA}, and Assumption~\ref{ASS:ideal} give the periodic-estimation error dynamics and aperiodic-estimation error dynamics as
\begin{align*}
	&\bm{e}_{\mathrm{p}}(t|t)=\bm{\epsilon}_{\mathrm{p}}(t-\Pi|t-\Pi) + \bm{S}_{\mathrm{p}}\bm{e}(t|t),\\
	&\bm{e}_{\mathrm{a}}(t|t)=\bm{\epsilon}_{\mathrm{a}}(t-\Pi|t-\Pi) + \bm{S}_{\mathrm{a}}\bm{e}(t|t),\\
	&\bm{\epsilon}_{\mathrm{p}}(t-\Pi|t-\Pi)=\\
	&\sum_{i=1}^{N}[\bm{G}_{\mathrm{p}i}\bm{e}_{\mathrm{p}}(t-i\Pi|t-i\Pi) + \bm{H}_{\mathrm{p}i}\bm{e}(t-i\Pi|t-i\Pi)],\\
	&\bm{\epsilon}_{\mathrm{a}}(t-\Pi|t-\Pi)=\\
	&\sum_{i=1}^{N}[\bm{G}_{\mathrm{a}i}\bm{e}_{\mathrm{a}}(t-i\Pi|t-i\Pi) + \bm{H}_{\mathrm{a}i}\bm{e}(t-i\Pi|t-i\Pi)].
\end{align*}
The expectation of the periodic-estimation error $\mathrm{E}[\bm{e}_{\mathrm{p}}(t|t)]$ and that of the aperiodic-estimation error $\mathrm{E}[\bm{e}_{\mathrm{a}}(t|t)]$ become zero as
\begin{align}
	&\mathrm{E}[\bm{\epsilon}_{\mathrm{p}}(t-\Pi|t-\Pi)]=\mathrm{E}[\bm{\epsilon}_{\mathrm{a}}(t-\Pi|t-\Pi)]=\bm{0}\notag\\
	&\hspace{1em}\land\mathrm{E}[\bm{e}(t|t)]=\bm{0}\Rightarrow
	\mathrm{E}[\bm{e}_{\mathrm{p}}(t|t)]=\mathrm{E}[\bm{e}_{\mathrm{a}}(t|t)]=\bm{0}.\notag
\end{align}
Using $\mathrm{E}[\bm{e}(t-1|t-1)]=\bm{0}\Rightarrow\mathrm{E}[\bm{e}(t|t)]=\bm{0}$ based on \eqref{eq:ub:ee},
\begin{align}
	&\mathrm{E}[\bm{\epsilon}_{\mathrm{p}}(t-\Pi|t-\Pi)]=\mathrm{E}[\bm{\epsilon}_{\mathrm{a}}(t-\Pi|t-\Pi)]=\bm{0}\notag\\
	&\hspace{1em}\land\mathrm{E}[\bm{e}(t-1|t-1)]=\bm{0}\Rightarrow
	\mathrm{E}[\bm{e}_{\mathrm{p}}(t|t)]=\mathrm{E}[\bm{e}_{\mathrm{a}}(t|t)]=\bm{0}.\notag
\end{align}
Additionally, the expectations of $\bm{\epsilon}_{\mathrm{p}}(t-\Pi|t-\Pi)$ and $\bm{\epsilon}_{\mathrm{a}}(t-\Pi|t-\Pi)$ become zero as
\begin{align}
	&\mathrm{E}[\bm{e}(t-1-n|t-1-n)]=\mathrm{E}[\bm{e}_{\mathrm{p}}(t-1-n|t-1-n)]\notag\\
	&=\mathrm{E}[\bm{e}_{\mathrm{a}}(t-1-n|t-1-n)]=\bm{0}\notag\\
	&\hspace{1em}\Rightarrow
	\mathrm{E}[\bm{\epsilon}_{\mathrm{p}}(t-\Pi|t-\Pi)]=\mathrm{E}[\bm{\epsilon}_{\mathrm{a}}(t-\Pi|t-\Pi)]=\bm{0}.\notag
\end{align}
These yield
\begin{align}
	&\mathrm{E}[\bm{e}(t-1-n|t-1-n)]=\mathrm{E}[\bm{e}_{\mathrm{p}}(t-1-n|t-1-n)]\notag\\
	&=\mathrm{E}[\bm{e}_{\mathrm{a}}(t-1-n|t-1-n)]=\bm{0}\notag\\
	&\hspace{1em}\Rightarrow\mathrm{E}[\bm{e}_{\mathrm{p}}(t|t)]=\mathrm{E}[\bm{e}_{\mathrm{a}}(t|t)]=\bm{0},\notag
\end{align}
which and $\mathrm{E}[\bm{e}(t-1|t-1)]=\bm{0}\Rightarrow\mathrm{E}[\bm{e}(t|t)]=\bm{0}$ derive
\begin{align}
	&\mathrm{E}[\bm{e}(-n|-n)]=\mathrm{E}[\bm{e}_{\mathrm{p}}(-n|-n)]=\mathrm{E}[\bm{e}_{\mathrm{a}}(-n|-n)]=\bm{0}\notag\\
	&\hspace{1em}\Rightarrow\mathrm{E}[\bm{e}(1|1)]=\mathrm{E}[\bm{e}_{\mathrm{p}}(1|1)]=\mathrm{E}[\bm{e}_{\mathrm{a}}(1|1)]=\bm{0},\notag\\
	&\hspace{1em}\Rightarrow\mathrm{E}[\bm{e}(2|2)]=\mathrm{E}[\bm{e}_{\mathrm{p}}(2|2)]=\mathrm{E}[\bm{e}_{\mathrm{a}}(2|2)]=\bm{0},\notag\\
	&\hspace{3em}\vdots\notag
\end{align}
Thus, the estimation of Algorithm~\ref{alg} is unbiased as
\begin{align}
	&\mathrm{E}[\bm{e}(-n|-n)]=\mathrm{E}[\bm{e}_{\mathrm{p}}(-n|-n)]=\mathrm{E}[\bm{e}_{\mathrm{a}}(-n|-n)]=\bm{0}\notag\\
	&\hspace{1em}\Rightarrow\ \forall t\in \mathbb{Z}_{>0},\ \mathrm{E}[\bm{e}(t|t)]=\mathrm{E}[\bm{e}_{\mathrm{p}}(t|t)]=\mathrm{E}[\bm{e}_{\mathrm{a}}(t|t)]=\bm{0}.\notag
\end{align}
\mbox{ }\vspace{-10mm}
\begin{flushright} $\blacksquare$ \end{flushright}

\begin{HMthm}\label{THM:ecm}\end{HMthm}\vspace{-3mm}
Assume Assumption~\ref{ASS:ideal} holds.
The error covariance matrix $\bm{P}(t|t)$ equals the sum of the periodic-error covariance matrix $\bm{P}_{\mathrm{p}}(t|t)$ and aperiodic-error covariance matrix $\bm{P}_{\mathrm{a}}(t|t)$
\begin{align}
	&\bm{P}(t|t)=\bm{P}_{\mathrm{p}}(t|t)+\bm{P}_{\mathrm{a}}(t|t),\
	\bm{P}_{\mathrm{p}}(t|t), \bm{P}_{\mathrm{a}}(t|t)\in \mathbb{R}^{n \times n}\notag\\
	&\bm{P}_{\mathrm{p}}(t|t)\coloneqq \mathrm{E}[\bm{e}_{\mathrm{p}}(t|t)\bm{e}_{\mathrm{p}}^{\mathrm{T}}(t|t)],\
	\bm{P}_{\mathrm{a}}(t|t)\coloneqq \mathrm{E}[\bm{e}_{\mathrm{a}}(t|t)\bm{e}_{\mathrm{a}}^{\mathrm{T}}(t|t)].\notag
\end{align}
$\bm{\mathrm{Proof.}}$
\begin{align}
	\bm{P}(t|t)&=\mathrm{E}[\bm{e}(t|t)\bm{e}^{\mathrm{T}}(t|t)],\notag\\
	&=\mathrm{E}[(\bm{e}_{\mathrm{p}}(t|t)+\bm{e}_{\mathrm{a}}(t|t))(\bm{e}_{\mathrm{p}}(t|t)+\bm{e}_{\mathrm{a}}(t|t))^{\mathrm{T}}].\notag
\end{align}
According to \eqref{eq:assume:ortho_PA_epea},
\begin{align}
	\bm{P}(t|t)&=\mathrm{E}[\bm{e}_{\mathrm{p}}(t|t)\bm{e}_{\mathrm{p}}^{\mathrm{T}}(t|t)]+\mathrm{E}[\bm{e}_{\mathrm{a}}(t|t)\bm{e}_{\mathrm{a}}^{\mathrm{T}}(t|t)],\notag\\
	&=\bm{P}_{\mathrm{p}}(t|t)+\bm{P}_{\mathrm{a}}(t|t).\notag
\end{align}
\mbox{ }\vspace{-10mm}
\begin{flushright} $\blacksquare$ \end{flushright}

%
The KF-PASF minimizes the sum of the periodic-error and aperiodic-error covariances $\bm{P}_{\mathrm{p}}(t|t)+\bm{P}_{\mathrm{a}}(t|t)$ by minimizing the error covariance $\bm{P}(t|t)$ with the algorithm of the Kalman filter.

\section{Examples} \label{sec:5}
This section shows estimation, comparison, and control examples.
The examples shown in Sections~\ref{sec:5-1}, \ref{sec:5-2}, and \ref{sec:5-4} used the system:
\begin{align*}
	&\bm{x}_{\mathrm{pa}}(t+1)=\bm{A}\bm{x}_{\mathrm{pa}}(t)+\bm{B}({u}(t)+v(t)),\\
	&{y}(t)=\bm{C}\bm{x}_{\mathrm{pa}}(t)+w(t),\\
	&\bm{x}_{\mathrm{pa}}(t)=\left[
	\begin{array}{cccc}
		x_1(t)&
		x_2(t)&
		x_3(t)
	\end{array}
	\right]^{\mathrm{T}},\
	\bm{B}=\left[
	\begin{array}{cccccc}
		0&
		0&
		1
	\end{array}
	\right]^{\mathrm{T}},\\
	&\bm{C}=\left[
	\begin{array}{cccccc}
		1&
		0&
		0
	\end{array}
	\right],\
	v\sim\mathcal{N}(0,10^{-4}),\
	w\sim\mathcal{N}(0,0.5).
\end{align*}
The PASF used by the examples was the first-order IIR ($N=1$), second-order IIR ($N=2$), third-order IIR ($N=3$), or 50th-order FIR periodic-pass and aperiodic-pass filters.
The parameters of the IIR filters were as follows
\begin{align}
	&T=1\ \mathrm{ms},\ \Pi=1000,\notag\\
	&\bm{P}(0|0)=\bm{0},\ \bm{Q}=\mathrm{diag}(0,\ 0,\ 10^{-8}),\ R=0.25,\notag\\
	&N=1:\notag\\
	&\begin{array}{lll}
		G_{\mathrm{p}1}=G_{\mathrm{a}1}=-\frac{\tilde{\rho} \Pi T-2}{\tilde{\rho} \Pi T+2},\
		H_{\mathrm{p}1}=\frac{\tilde{\rho} \Pi T}{\tilde{\rho} \Pi T+2},\
		H_{\mathrm{a}1}=-\frac{2}{\tilde{\rho} \Pi T+2},\\
		S_{\mathrm{p}}=\frac{\tilde{\rho} \Pi T}{\tilde{\rho} \Pi T+2},\
		S_{\mathrm{a}}=\frac{2}{\tilde{\rho} \Pi T+2},
	\end{array}\notag\\
	&N=2:\notag\\
	&\begin{array}{ll}
		G_{\mathrm{p}1}=G_{\mathrm{a}1}=-2\frac{\tilde{\rho} \Pi T-2}{\tilde{\rho} \Pi T+2},\
		G_{\mathrm{p}2}=G_{\mathrm{a}2}=-\frac{(\tilde{\rho} \Pi T-2)^2}{(\tilde{\rho} \Pi T+2)^2},\\
		H_{\mathrm{p}1}=2\frac{(\tilde{\rho} \Pi T)^2}{(\tilde{\rho} \Pi T+2)^2},\
		H_{\mathrm{p}2}=\frac{(\tilde{\rho} \Pi T)^2}{(\tilde{\rho} \Pi T+2)^2},\\
		H_{\mathrm{a}1}=-\frac{8}{(\tilde{\rho} \Pi T+2)^2},\
		H_{\mathrm{a}2}=\frac{4}{(\tilde{\rho} \Pi T+2)^2},\\
		S_{\mathrm{p}}=\frac{(\tilde{\rho} \Pi T)^2}{(\tilde{\rho} \Pi T+2)^2},\
		S_{\mathrm{a}}=\frac{4}{(\tilde{\rho} \Pi T+2)^2},
	\end{array}\notag\\
	&N=3:\notag\\
	&\begin{array}{ll}
		G_{\mathrm{p}1}=G_{\mathrm{a}1}=-3\frac{\tilde{\rho} \Pi T-2}{\tilde{\rho} \Pi T+2},\
		G_{\mathrm{p}2}=G_{\mathrm{a}2}=-3\frac{(\tilde{\rho} \Pi T-2)^2}{(\tilde{\rho} \Pi T+2)^2},\\
		G_{\mathrm{p}3}=G_{\mathrm{a}3}=-\frac{(\tilde{\rho} \Pi T-2)^3}{(\tilde{\rho} \Pi T+2)^3},
	\end{array}\notag\\
	&\begin{array}{lll}
		H_{\mathrm{p}1}=3\frac{(\tilde{\rho} \Pi T)^3}{(\tilde{\rho} \Pi T+2)^3},&
		H_{\mathrm{p}2}=3\frac{(\tilde{\rho} \Pi T)^3}{(\tilde{\rho} \Pi T+2)^3},&
		H_{\mathrm{p}3}=\frac{(\tilde{\rho} \Pi T)^3}{(\tilde{\rho} \Pi T+2)^3},\\
		H_{\mathrm{a}1}=-\frac{24}{(\tilde{\rho} \Pi T+2)^3},&
		H_{\mathrm{a}2}=\frac{24}{(\tilde{\rho} \Pi T+2)^3},&
		H_{\mathrm{a}3}=-\frac{8}{(\tilde{\rho} \Pi T+2)^3},
	\end{array}\notag\\
	&\begin{array}{ll}
		S_{\mathrm{p}}=\frac{(\tilde{\rho} \Pi T)^3}{(\tilde{\rho} \Pi T+2)^3},\
		S_{\mathrm{a}}=\frac{8}{(\tilde{\rho} \Pi T+2)^3}.
	\end{array}\notag
\end{align}
The 50th-order FIR filter was designed using the MATLAB function \textit{firceqrip()}.
\begin{figure*}[t!]
	\begin{center}
		\includegraphics[width=0.9\hsize]{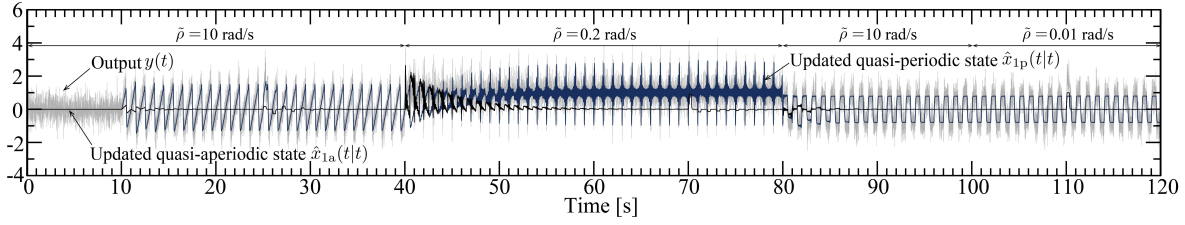}
	\end{center}
	\caption{Estimation of the quasi-periodic and quasi-aperiodic states with the variations in the input $u(t)$ and the separation frequency $\tilde{\rho}$.}\label{fig:ex:ex}
\end{figure*}

\begin{figure*}[t!]
	\begin{minipage}{0.33\hsize}
		\begin{center}
			\includegraphics[width=0.97\hsize]{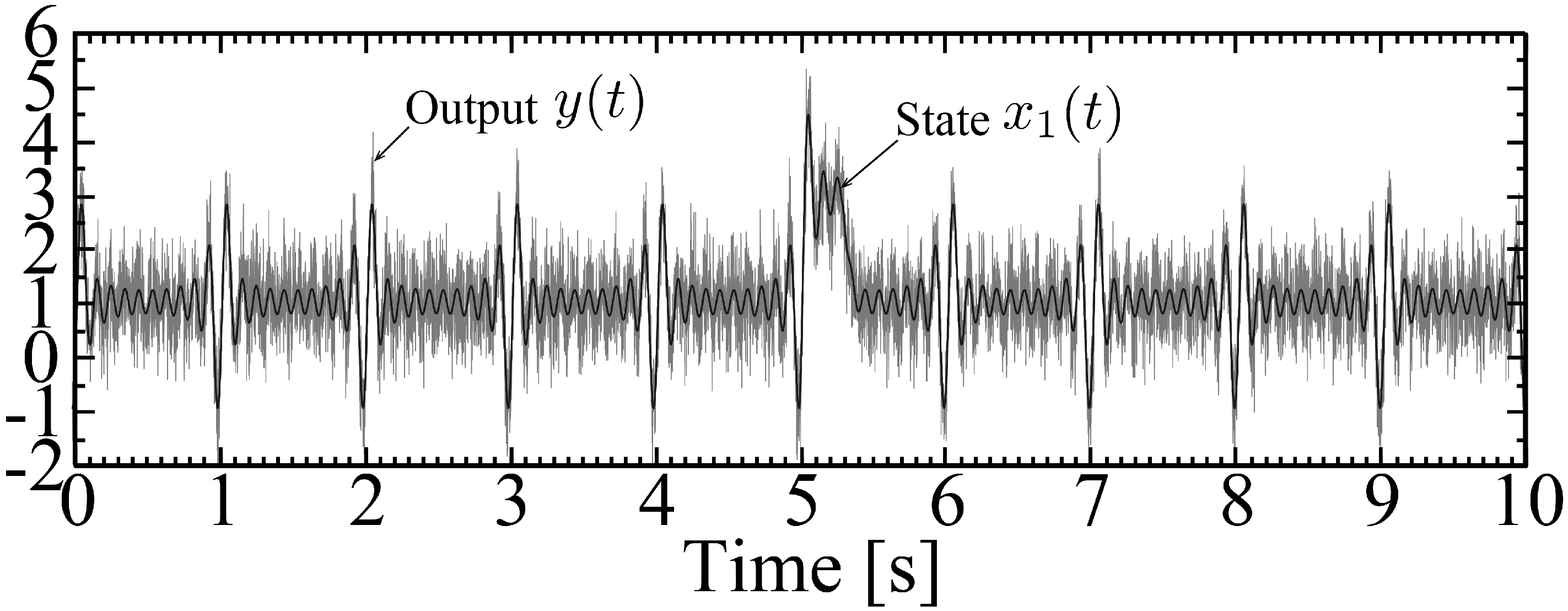}\\
			(a) Output and state.
		\end{center}
	\end{minipage}
	\begin{minipage}{0.33\hsize}
		\begin{center}
			\includegraphics[width=0.97\hsize]{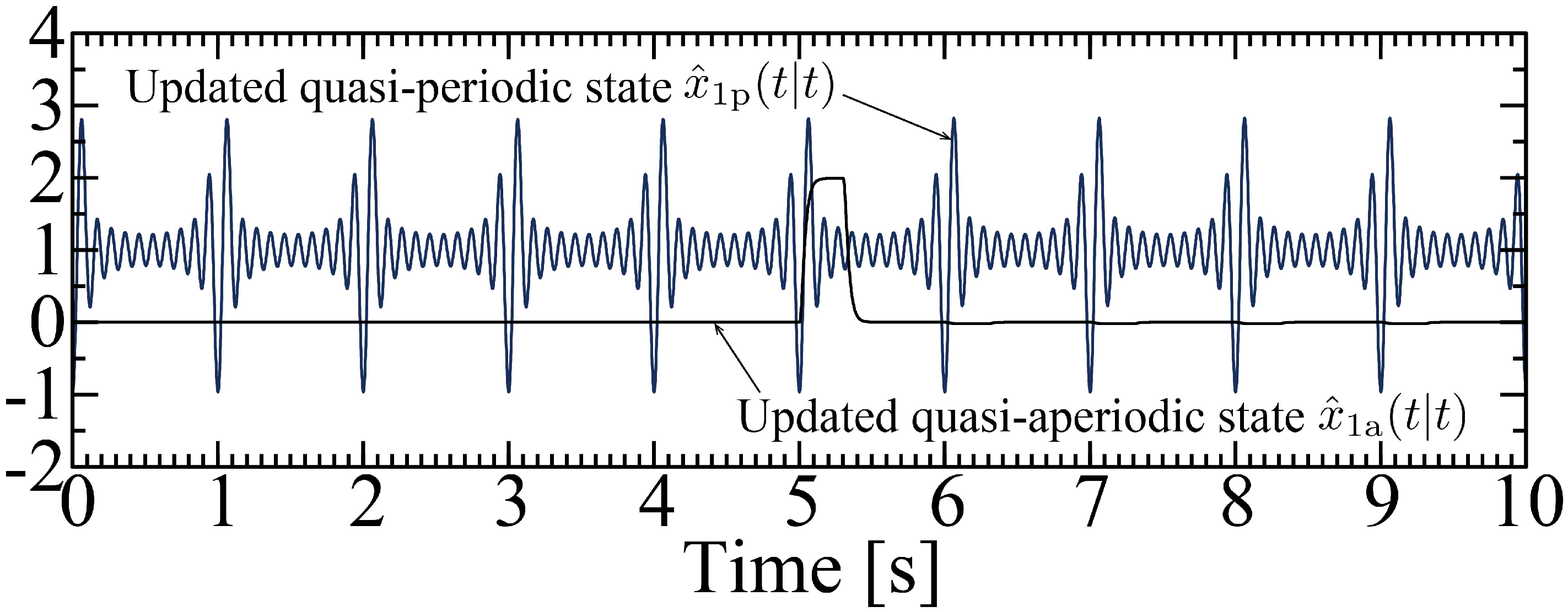}\\
			(b) First-order IIR filter.
		\end{center}
	\end{minipage}
	\begin{minipage}{0.33\hsize}
		\begin{center}
			\includegraphics[width=0.97\hsize]{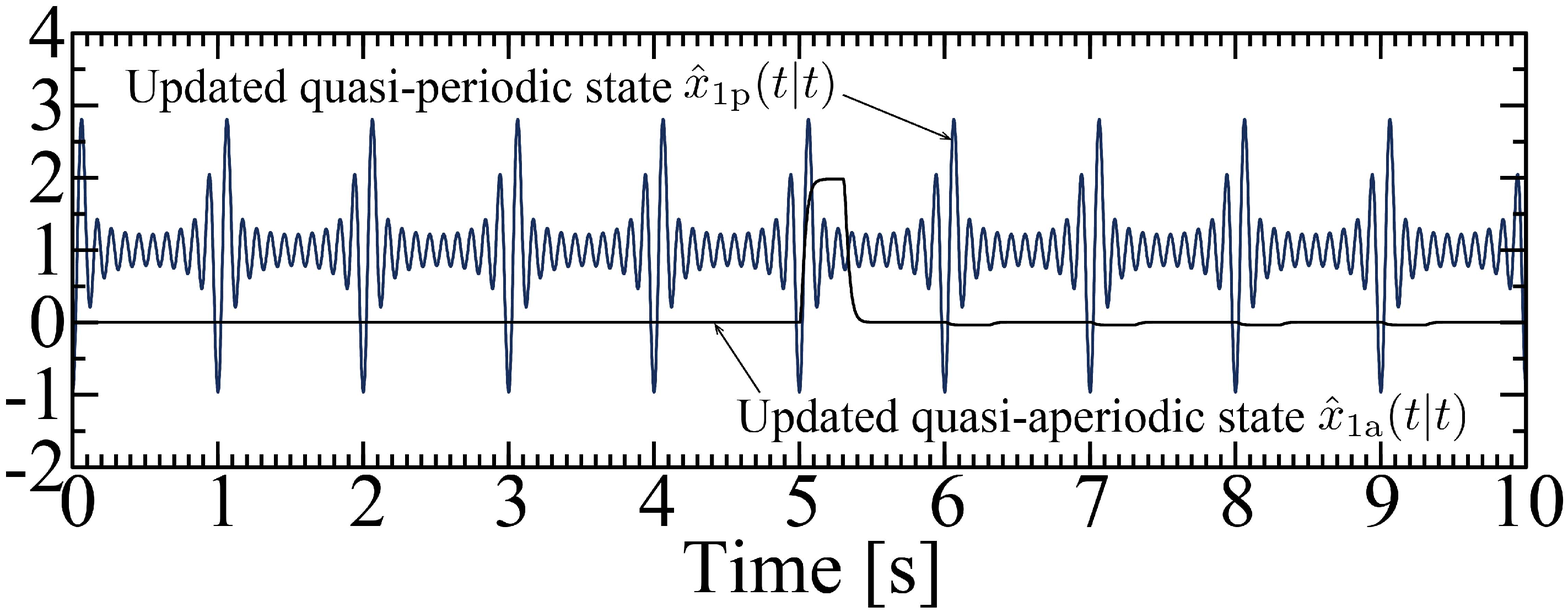}\\
			(c) Second-order IIR filter.
		\end{center}
	\end{minipage}
	\begin{minipage}{0.33\hsize}
		\begin{center}
			\includegraphics[width=0.97\hsize]{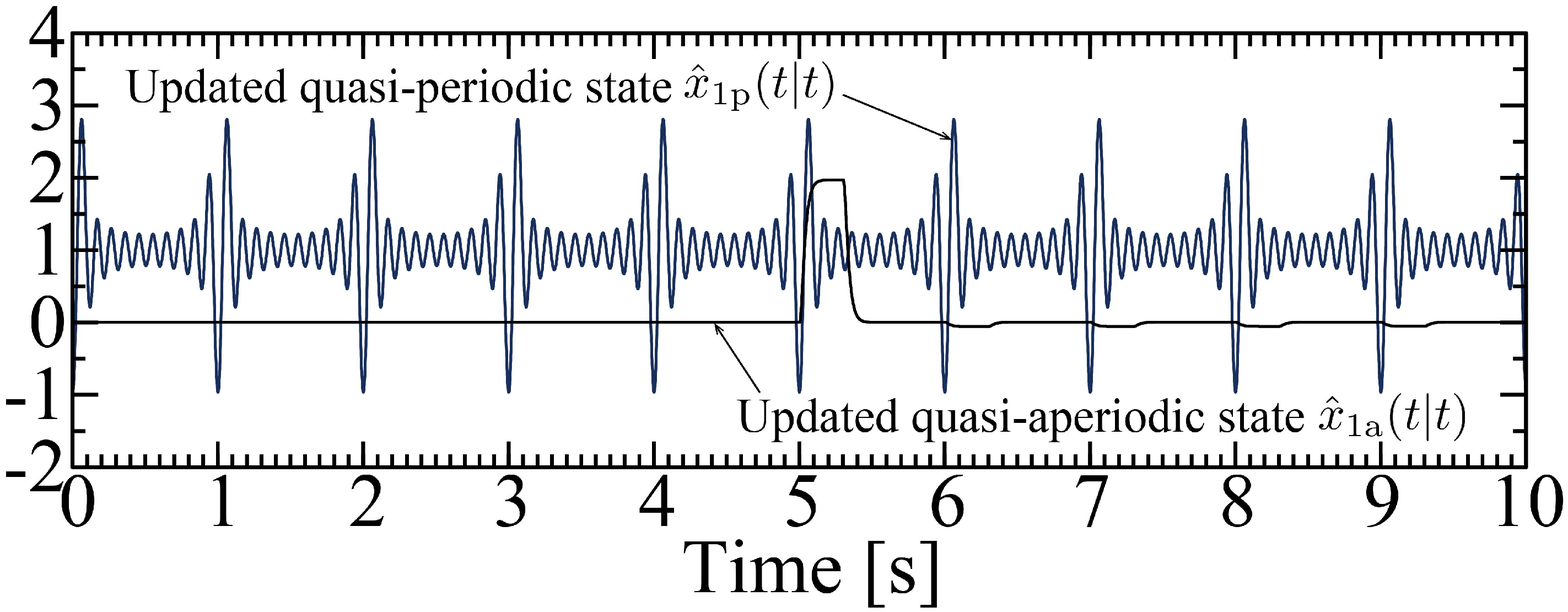}\\
			(d) Third-order IIR filter.
		\end{center}
	\end{minipage}
	\begin{minipage}{0.33\hsize}
		\begin{center}
			\includegraphics[width=0.97\hsize]{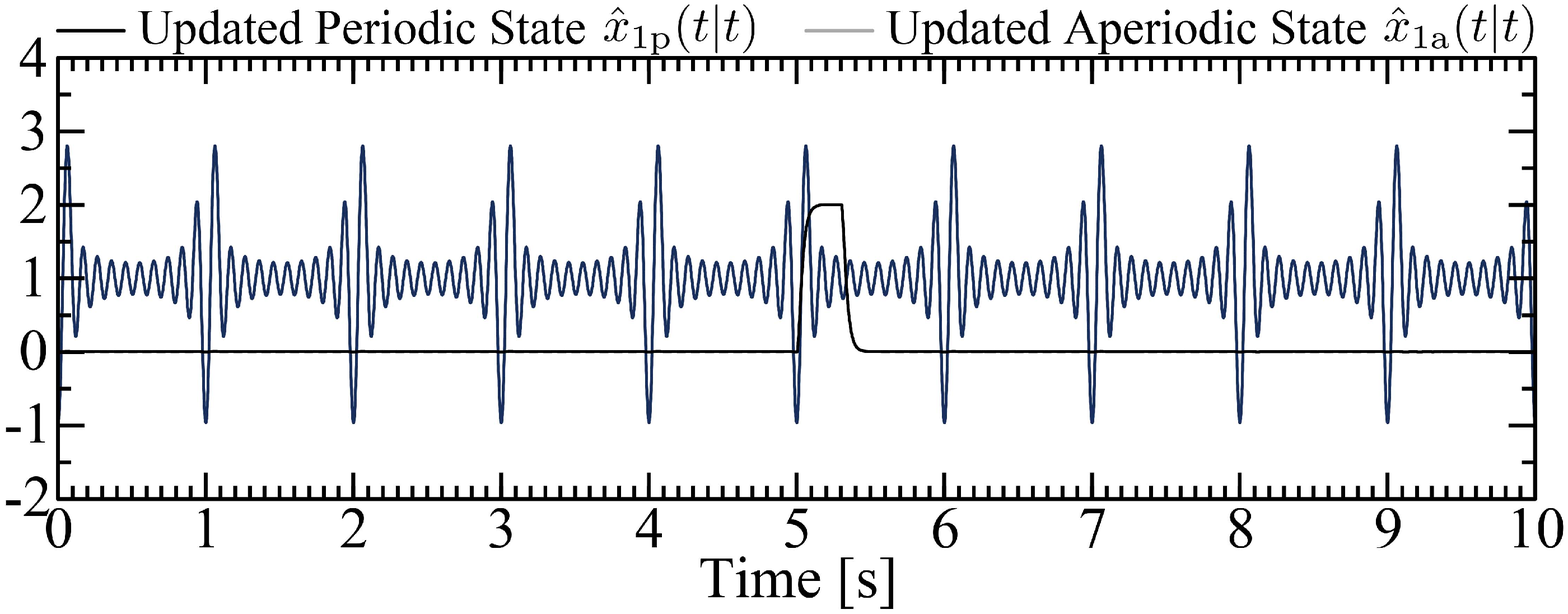}\\
			(e) Fiftieth-order FIR filter.
		\end{center}
	\end{minipage}
	\begin{minipage}{0.33\hsize}
		\begin{center}
			\includegraphics[width=0.97\hsize]{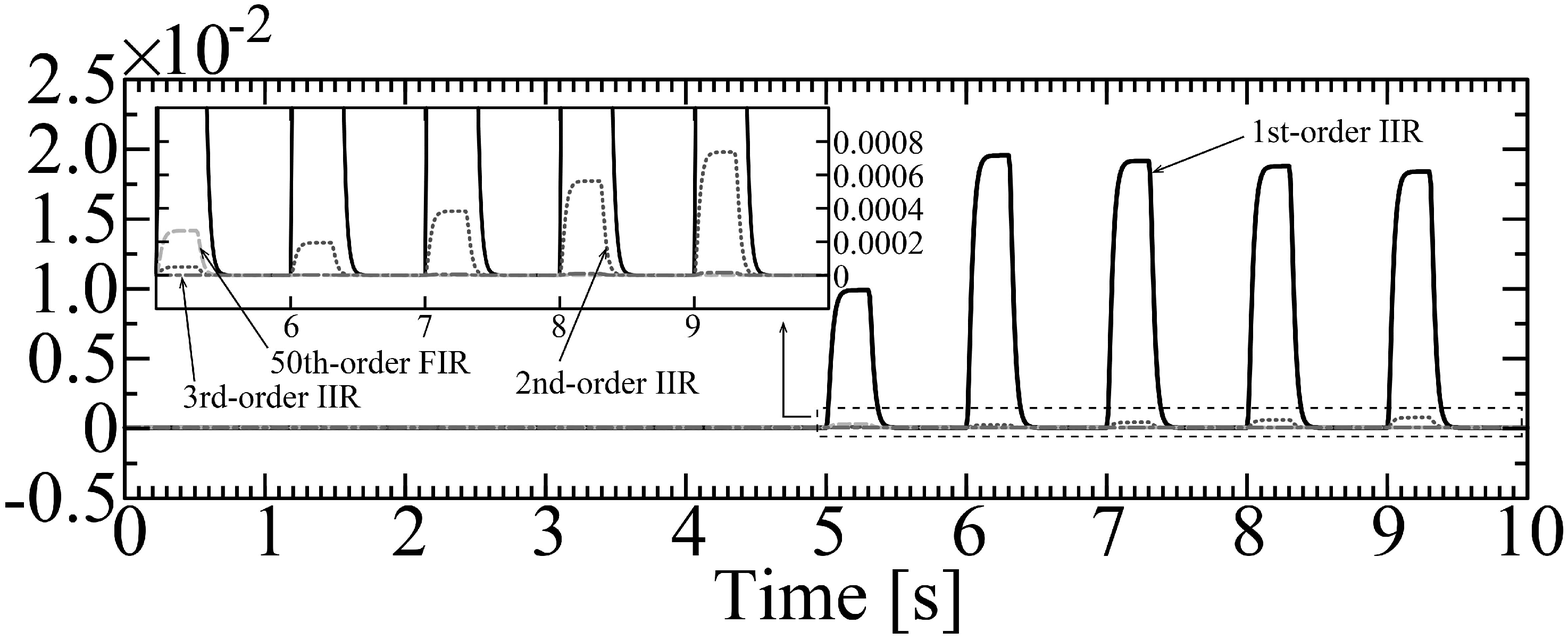}\\
			(f) Interferences.
		\end{center}
	\end{minipage}
	\caption{Four estimation results of the four periodic-pass and aperiodic-pass filters.}\label{fig:ex:est}
\end{figure*}

\subsection{Estimation of Quasi-Periodic and Quasi-Aperiodic States with Separation Frequency Change} \label{sec:5-1}
In this example, the KF-PASF in Algorithm~\ref{alg}, using the first-order IIR periodic-pass and aperiodic-pass filters, estimated the quasi-periodic and quasi-aperiodic states.
The example employed the matrix, signals, and separation frequency
\begin{align}
	\bm{A}&=\left[
	\begin{array}{cccccc}
		1&T&0\\
		0&1&T\\
		-2500&-100&0
	\end{array}
	\right],\
	u(t)=2500(u_{1}(t)+u_{2}(t)),\notag\\
	u_{1}(t)&=\left\{
	\begin{array}{cl}
		0&\mathrm{if}\ Tt<10\ \mathrm{s}\\
		\displaystyle\sum_{i=1}^{10}\dfrac{(-1)^i}{-i}\sin(i 2\pi Tt)&\mathrm{if}\ 10\ \mathrm{s}\leq Tt<40\ \mathrm{s}\\
		\displaystyle\sum_{i=1}^{10}i^20.01\sin(i 2\pi Tt)&\mathrm{if}\ 40\ \mathrm{s}\leq Tt<80\ \mathrm{s}\\
		\multicolumn{2}{c}{\displaystyle\sum_{i=1}^{10}\dfrac{1}{(2i-1)}\sin((2i-1) 2\pi Tt)\ \mathrm{otherwise}}
	\end{array}
	\right.,\notag\\
	u_{2}(t)&=\left\{
	\begin{array}{cl}
		1&\mathrm{if}\ 25\ \mathrm{s}\leq Tt\leq 25.3\ \mathrm{s}\\
		&\hspace{1em}\lor70\ \mathrm{s}\leq Tt\leq 70.3\ \mathrm{s}\\
		&\hspace{1em}\lor110\ \mathrm{s}\leq Tt\leq 110.3\ \mathrm{s}\\
		0&\mathrm{otherwise}
	\end{array}
	\right.,\notag\\
	\tilde{\rho}&=\left\{
	\begin{array}{cl}
		10\ \mathrm{rad/s}&\mathrm{if}\ Tt<40\ \mathrm{s}\lor 80\ \mathrm{s}\leq Tt\leq 100\ \mathrm{s}\\
		0.2\ \mathrm{rad/s}&\mathrm{if}\ 40\ \mathrm{s}\leq Tt\leq 80\ \mathrm{s}\\
		0.01\ \mathrm{rad/s}&\mathrm{otherwise}
	\end{array}
	\right..\notag
\end{align}
Fig.~\ref{fig:ex:ex} shows the estimation result for $\hat{x}_{1\mathrm{p}}(t|t)$ and $\hat{x}_{1\mathrm{a}}(t|t)$, where the KF-PASF estimated the three different quasi-periodic states with the three separation frequencies.
The large and small separation frequencies provided fast and slow convergences of the updated quasi-periodic and quasi-aperiodic states, respectively.
Meanwhile, the large separation frequency provided the states that were much affected by $u_2(t)$ even while $u_2(t)$ became zero from one, and the small separation frequency provided the rigid separation that was not much affected by $u_2(t)$ while $u_2(t)=0$.

\subsection{Comparison of Realizations} \label{sec:5-2}
This example compared the KF-PASFs based on the different realizations of the three IIR filters and an FIR filter with the matrix, signals, and the separation frequency
\begin{align}
	\bm{A}&=\left[
	\begin{array}{cccccc}
		1&T&0\\
		0&1&T\\
		-2500&-100&0
	\end{array}
	\right],\
	u(t)=2500(u_{1}(t)+u_{2}(t)),\notag\\
	u_{1}(t)&=1+\sum_{i=1}^{10}i^20.01\sin(i2\pi Tt),\notag\\
	u_{2}(t)&=\left\{
	\begin{array}{cl}
		 2&\mathrm{if}\ 15 < t \leq 15.3\\
		 0&\mathrm{otherwise}
	\end{array}
	\right.,\
	\tilde{\rho}=0.01\ \mathrm{rad/s}.\notag
\end{align}
Their initial updated states $\hat{\bm{x}}_{\mathrm{pa}}(-n|-n)$, $\hat{\bm{x}}_{\mathrm{p}}(-n|-n)$, and $\hat{\bm{x}}_{\mathrm{a}}(-n|-n)$ were given beforehand.
Fig.~\ref{fig:ex:est} shows the estimation results for $\hat{x}_{1\mathrm{p}}(t|t)$ and $\hat{x}_{1\mathrm{a}}(t|t)$, where the output and actual state are depicted in (a) and the estimated states are depicted in (b)--(e).
According to the estimated states, there was no significant difference among the four filters; however, the updated quasi-aperiodic states given by the IIR filters were slightly oscillatory in contrast to the state given by the FIR filter.
Furthermore, (f) shows the four interferences that were the separated quasi-aperiodic states from the updated quasi-periodic states $\hat{x}_{1 \mathrm{p}}(t|t)$.
The increase in the order of the IIR filter reduced the interference, whereas the interference of the 50th-order FIR filter was larger than that of the third-order IIR filter.

\begin{figure*}[t!]
	\begin{minipage}{0.33\hsize}
		\begin{center}
			\includegraphics[width=0.97\hsize]{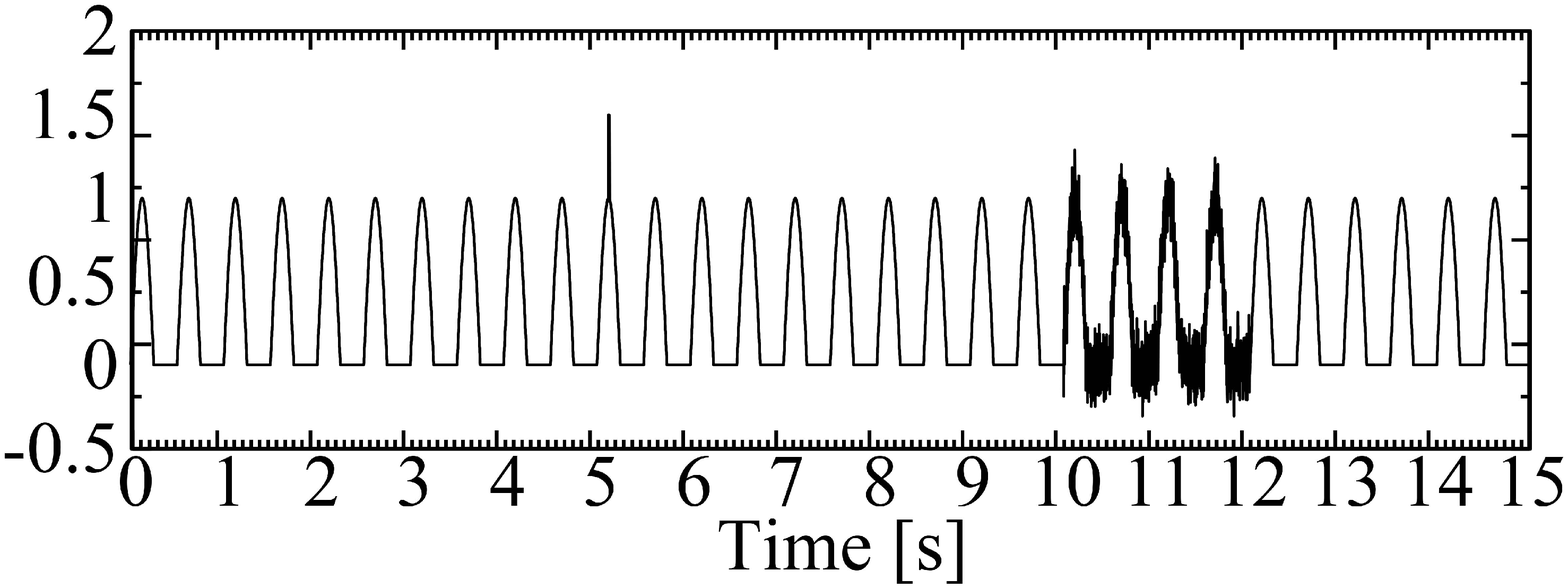}
		\end{center}
	\end{minipage}
	\begin{minipage}{0.33\hsize}
		\begin{center}
			\includegraphics[width=0.97\hsize]{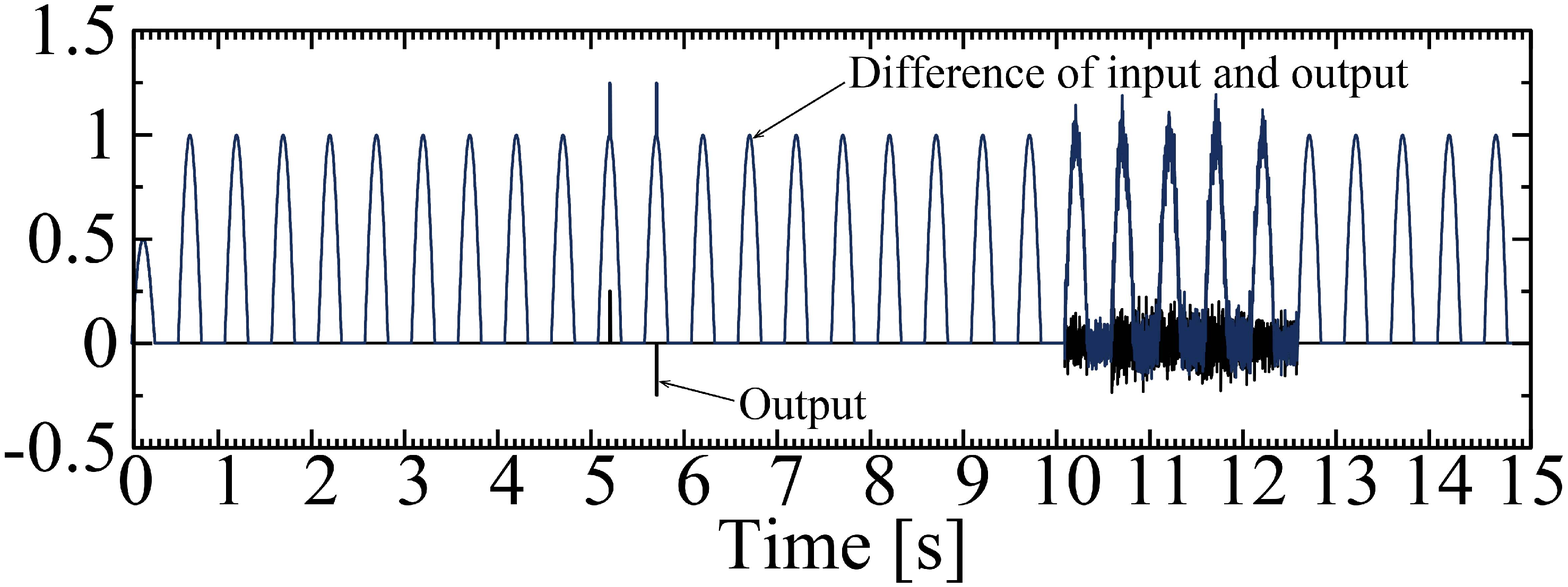}
		\end{center}
	\end{minipage}
	\begin{minipage}{0.33\hsize}
		\begin{center}
			\includegraphics[width=0.97\hsize]{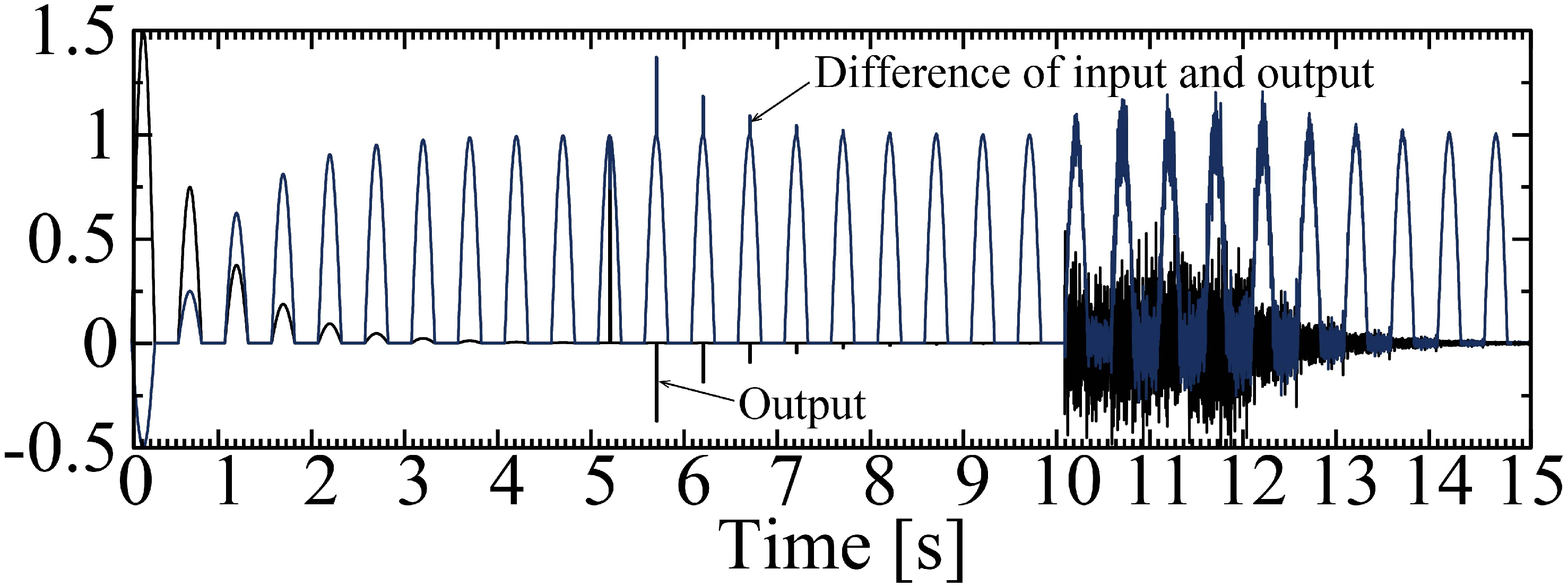}
		\end{center}
	\end{minipage}
	\begin{minipage}{0.33\hsize}
		\begin{center}
			(a) Input periodic/aperiodic signal.
		\end{center}
	\end{minipage}
	\begin{minipage}{0.33\hsize}
		\begin{center}
			(b) Comb filter $C_{\mathrm{omb1}}(z^{-1})$.
		\end{center}
	\end{minipage}
	\begin{minipage}{0.33\hsize}
		\begin{center}
			(c) Comb filter $C_{\mathrm{omb2}}(z^{-1})$.
		\end{center}
	\end{minipage}
	\begin{minipage}{0.33\hsize}
		\begin{center}
			\includegraphics[width=0.97\hsize]{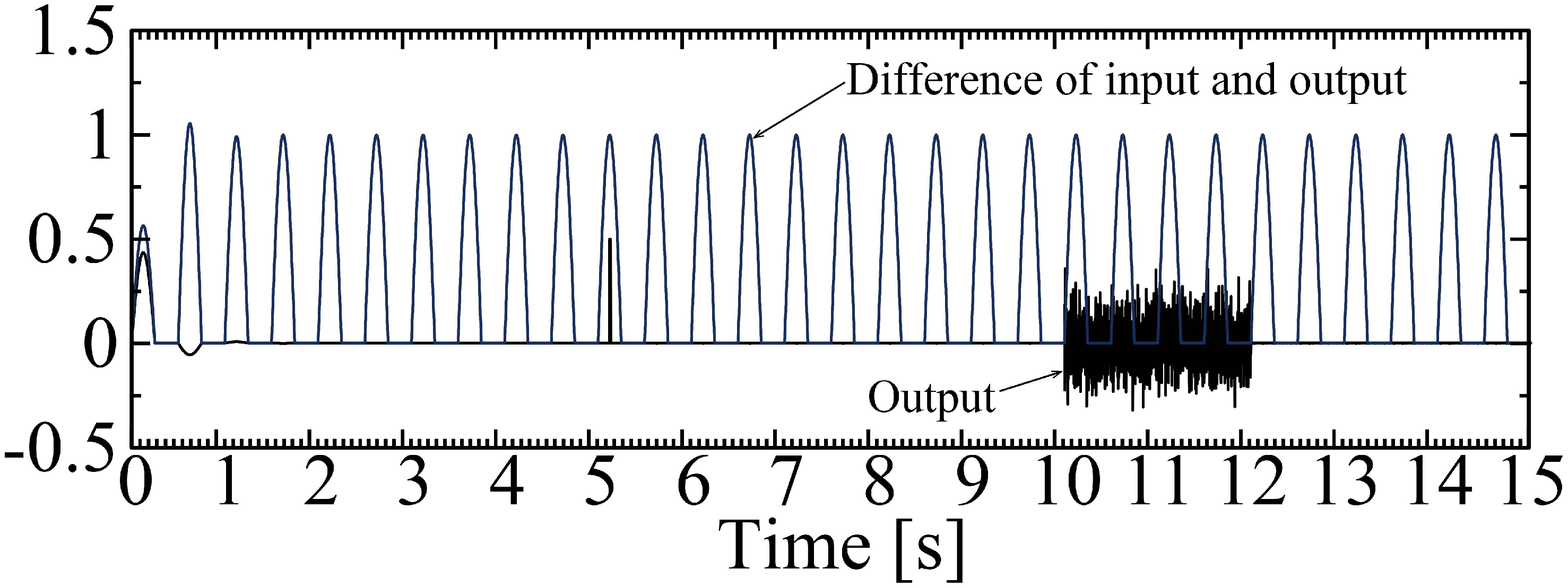}
		\end{center}
	\end{minipage}
	\begin{minipage}{0.33\hsize}
		\begin{center}
			\includegraphics[width=0.97\hsize]{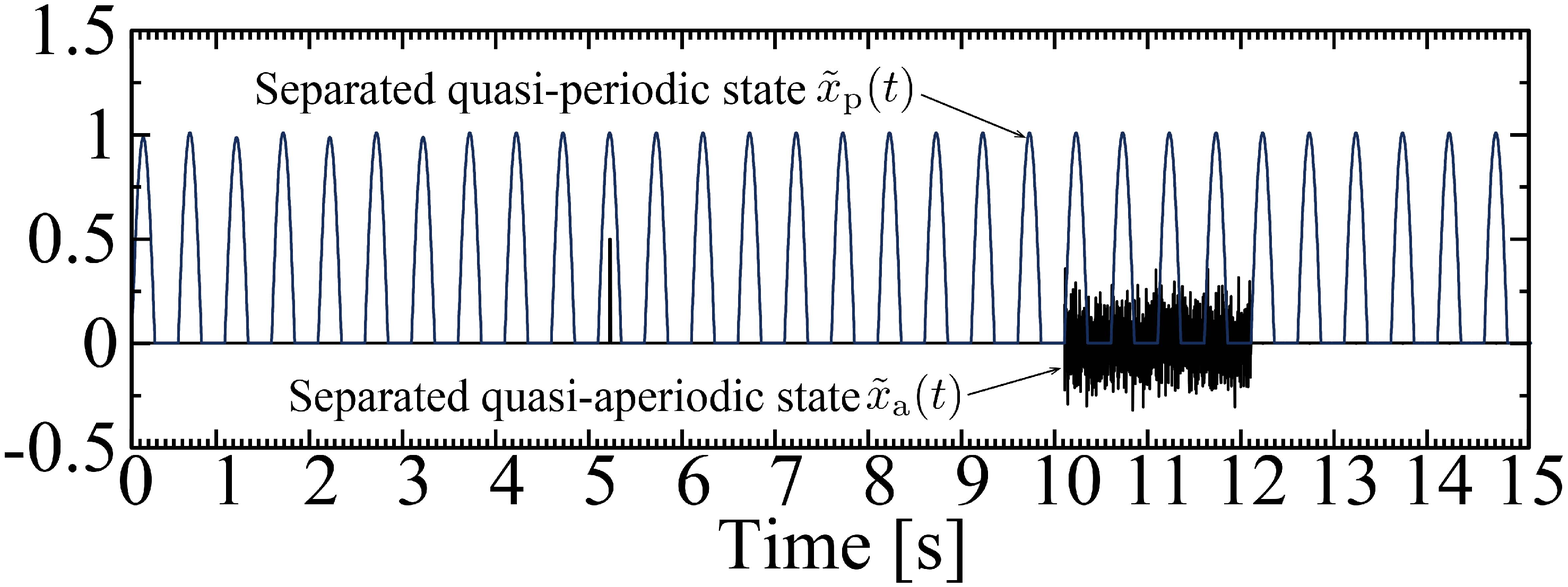}
		\end{center}
	\end{minipage}
	\begin{minipage}{0.33\hsize}
		\begin{center}
			\includegraphics[width=0.97\hsize]{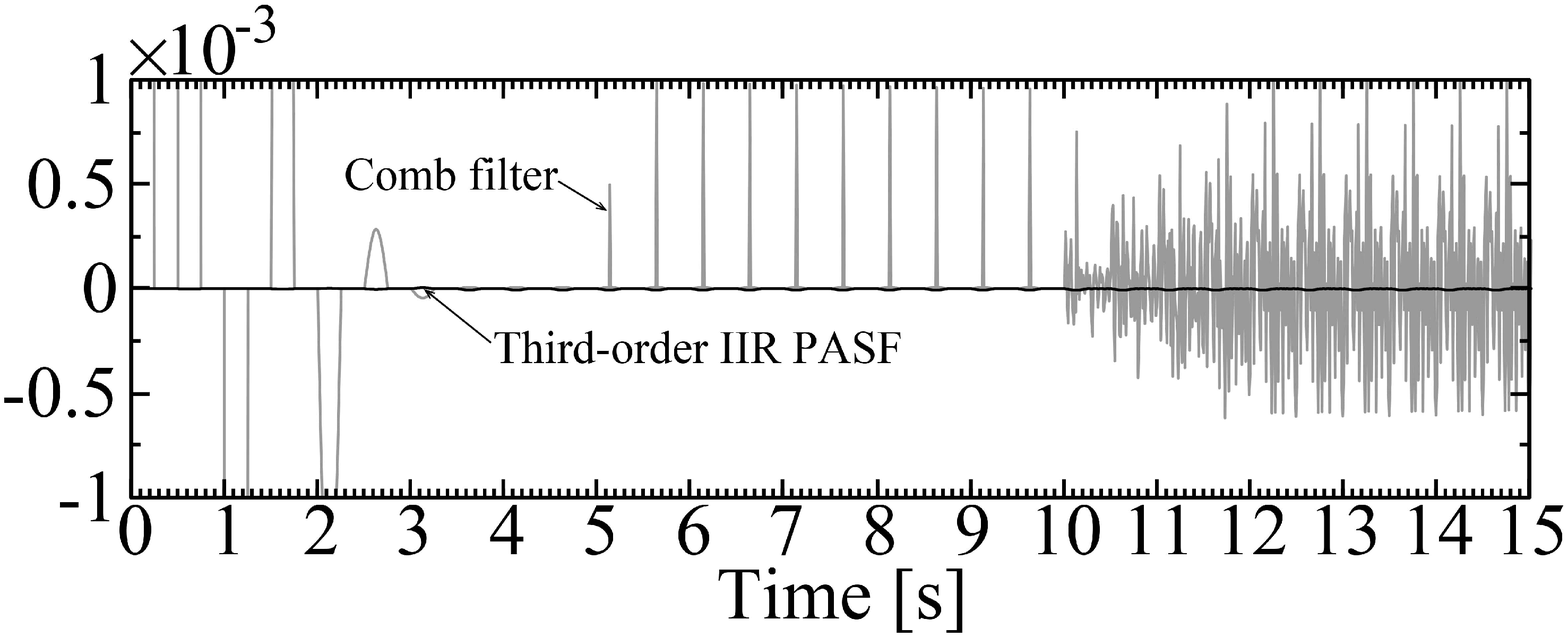}
		\end{center}
	\end{minipage}
	\begin{minipage}{0.33\hsize}
		\begin{center}
			(d) Comb filter $C_{\mathrm{omb3}}(z^{-1})$.\\\mbox{ }
		\end{center}
	\end{minipage}
	\begin{minipage}{0.33\hsize}
		\begin{center}
			(e) Third-order IIR PASF.\\\mbox{ }
		\end{center}
	\end{minipage}
	\begin{minipage}{0.33\hsize}
		\begin{center}
			(f) Interferences of the third-order IIR PASF and comb filter $C_{\mathrm{omb3}}(z^{-1})$.
		\end{center}
	\end{minipage}
	\caption{Comparison of the third-order PASF with comb filters.}\label{fig:ex:comb}
\end{figure*}

\subsection{Comparison of PASF with Comb Filters} \label{sec:5-3}
This example compared the third-order IIR PASF with three conventional comb filters.
The input periodic/aperiodic signal $x_{\mathrm{pa}}$ was
\begin{align*}
	x_{\mathrm{pa}}(t)&=x_{\mathrm{p}}(t)+x_{\mathrm{a}}(t),\
	\nu\sim\mathcal{N}(0,0.01),\\
	x_{\mathrm{p}}(t)&=\left\{
	\begin{array}{cl}
		\sin(4\pi Tt)&\mathrm{if}\ (t\ \mathrm{mod}\ 500)<250\\
		0&\mathrm{otherwise}
	\end{array}
	\right. ,\\
	x_{\mathrm{a}}(t)&=\left\{
	\begin{array}{cl}
		0.5&\mathrm{if}\ 5.125\ \mathrm{s}<Tt\leq 5.135\ \mathrm{s}\\
		\nu&\mathrm{if}\ 10\ \mathrm{s}<Tt\leq 12\ \mathrm{s}\\
		0&\mathrm{otherwise}
	\end{array}
	\right.,
\end{align*}
and the separation frequency $\tilde{\rho}$ of the third-order PASF was
\begin{align}
	\label{eq:}
	\tilde{\rho}=\left\{
	\begin{array}{cl}
		1000\ \mathrm{rad/s}&\mathrm{if}\ Tt< 4\ \mathrm{s}\\
		0.001\ \mathrm{rad/s}&\mathrm{otherwise}
	\end{array}
	\right..\notag
\end{align}

The comparison used the comb filter presented by \cite{2013_Sugiura_Comb}
\begin{align*}
	C_{\mathrm{omb}}(z^{-1})\coloneqq 1-\frac{(1-g)(1-b)}{2}\frac{1+z^{-\Pi}}{1-bz^{-\Pi}},
\end{align*}
which was designed as $C_{\mathrm{omb1}}(z^{-1})$ with $b=0$ and $g=0$ and $C_{\mathrm{omb2}}(z^{-1})$ with $b=0.5$ and $g=0$
\begin{align*}
	C_{\mathrm{omb1}}(z^{-1})\coloneqq \frac{1-z^{-\Pi}}{2},\ C_{\mathrm{omb2}}(z^{-1})=\frac{3}{2}\frac{1-z^{-\Pi}}{2-z^{-\Pi}}.
\end{align*}
They are equivalent to the feedforward and feedback comb filters of \cite{2017_Liu_Comb}, respectively.
Additionally, the comparison employed the comb filter presented by \cite{2018_aslan}
\begin{align*}
	&C_{\mathrm{omb3}}(z^{-1})\coloneqq \frac{\beta[1-z^{-\Pi}]}{1-\alpha z^{-\Pi}},\\
	&\alpha\coloneqq \frac{1-\gamma}{1+\gamma},\
	\beta\coloneqq \frac{1}{1+\gamma},\
	\gamma\coloneqq \frac{\sqrt{1-|G_{\mathrm{cb}}|^2}}{|G_{\mathrm{cb}}|}\tan{\frac{\pi}{2Q}},
\end{align*}
where $|G_{\mathrm{cb}}|=0.708$ and
\begin{align*}
	Q= \left\{
	\begin{array}{cl}
		1.717&\mathrm{if}\ Tt< 4\ \mathrm{s}\\
		1591&\mathrm{otherwise}
	\end{array}
	\right..
\end{align*}
The comb filters are filters that eliminate harmonics; hence, this example regarded the outputs of the comb filters as quasi-aperiodic signals for the comparison.
Furthermore, the difference between the input periodic/aperiodic signal and output was accordingly regarded as a quasi-periodic signal.
The periodic-pass and aperiodic-pass filters of the third-order IIR PASF were $F_{\mathrm{p}}(z^{-1})$ and $F_{\mathrm{a}}(z^{-1})$ with $N=3$ of \eqref{eq:IIR:LowHigh:z1}, respectively.
All initial states of the PASF and comb filters were set to zero.
Fig.~\ref{fig:ex:comb} shows the comparative results.
Fig.~\ref{fig:ex:comb}(a) depicts the input periodic/aperiodic signal $x_{\mathrm{pa}}$, and the separated signals are depicted in (b)--(e).
According to (b), (c), and (e), the third-order IIR PASF realized a more rigid separation than the comb filters $C_{\mathrm{omb1}}(z^{-1})$ and $C_{\mathrm{omb2}}(z^{-1})$.
According to (d) and (e), the separation results of the third-order IIR PASF and the comb filter $C_{\mathrm{omb3}}(z^{-1})$ were similar, but the interference, which was the separated quasi-aperiodic signal from the separated quasi-periodic signal, of the PASF was smaller than that of the comb filter, as shown in (f).

\subsection{Periodic/Aperiodic Separation Control Based on KF-PASF} \label{sec:5-4}
\begin{figure}[t!]
	\begin{center}
		\includegraphics[width=0.9\hsize]{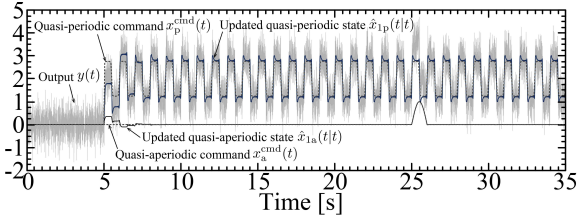}
	\end{center}
	\caption{Periodic/Aperiodic separation control. The control was activated at 5 s.}\label{fig:ex:ctrl}
\end{figure}

In this control example, the KF-PASF, using the first-order IIR periodic-pass and aperiodic-pass filters, was applied to the periodic/aperiodic separation control, which is the control of the quasi-periodic and quasi-aperiodic states.
This control example employed
\begin{align}
	&\bm{A}=\left[
	\begin{array}{cccccc}
		1&T&0\\
		0&1&T\\
		0&0&0
	\end{array}
	\right],\
	\tilde{\rho}=\left\{
	\begin{array}{cl}
		10\ \mathrm{rad/s}&\mathrm{if}\ Tt<20\ \mathrm{s}\\
		0.01\ \mathrm{rad/s}&\mathrm{otherwise}
	\end{array}
	\right.,\notag\\
	&x_{\mathrm{p}}^{\mathrm{cmd}}(t) \coloneqq  2 + \sum_{i=1}^{10}\dfrac{1}{(2i-1)}\sin((2i-1) 2\pi Tt),\notag\\
	&x_{\mathrm{a}}^{\mathrm{cmd}}(t) \coloneqq  \left\{
	\begin{array}{cl}
		\sin(\pi(Tt-25)) & \mathrm{if}\ 25\ \mathrm{s}< Tt \leq 26\ \mathrm{s}\\
		0&\mathrm{otherwise}
	\end{array}
	\right.,	\notag\\
	&u(t) =  \left\{
	\begin{array}{cl}
		0 & \mathrm{if}\ Tt < 5\ \mathrm{s}\\
		u_{\mathrm{p}}(t)+u_{\mathrm{a}}(t)&\mathrm{otherwise}
	\end{array}
	\right.,\notag\\
	&u_{\mathrm{p}}(t)\coloneqq 900(x_{\mathrm{p}}^{\mathrm{cmd}}(t)-\hat{x}_{1\mathrm{p}}(t|t)) + 60(\dot{x}_{\mathrm{p}}^{\mathrm{cmd}}(t)-\hat{x}_{2\mathrm{p}}(t|t)),\notag\\
	&u_{\mathrm{a}}(t)\coloneqq 2500(x_{\mathrm{a}}^{\mathrm{cmd}}(t)-\hat{x}_{1\mathrm{a}}(t|t)) + 100(\dot{x}_{\mathrm{a}}^{\mathrm{cmd}}(t)-\hat{x}_{2\mathrm{a}}(t|t)).\notag
\end{align}
Fig.~\ref{fig:ex:ctrl} shows the control results for $\hat{x}_{1\mathrm{p}}(t|t)$ and $\hat{x}_{1\mathrm{a}}(t|t)$, where the quasi-periodic and quasi-aperiodic states converged at their commands under the noise with different feedback gains for the states.

\section{Conclusion} \label{sec:6}
This paper defined periodic/aperiodic state composed of orthogonal quasi-periodic and quasi-aperiodic states, which further defined linear periodic-pass and aperiodic-pass functions.
It was demonstrated that the sum of the quasi-periodic (quasi-aperiodic) states or zero is quasi-periodic (quasi-aperiodic) or zero.
Similarly, the product of any value and the quasi-periodic state (quasi-aperiodic) or zero is quasi-periodic (quasi-aperiodic) or zero.
Moreover, based on the definitions, the functions were realized as causal linear periodic-pass and aperiodic-pass filters, which represent the PASF.
The realized high-order IIR and FIR periodic-pass and aperiodic-pass filters enhanced the slope of the band-stop characteristics and reduced the interference between the separated quasi-periodic and quasi-aperiodic states.
Additionally, the complementary realization eliminated the phase lag of the aperiodic-pass filter at the band-pass frequencies.
Lastly, the KF-PASF that integrates the PASF and Kalman filter achieved the unbiased estimation of the quasi-periodic and quasi-aperiodic states with the minimum sum of the periodic-error and aperiodic-error covariances.
A limitation of this study is that the realization error between the periodic-pass and aperiodic-pass functions and the causal periodic-pass and aperiodic-pass filters is inevitable.
Hence, in practical use, the quasi-periodic and quasi-aperiodic states are only almost separated and Assumption~\ref{ASS:ideal} is only almost satisfied.
Nevertheless, this imperfect realization is similar to the imperfect realization of low-pass and high-pass filters; therefore, the PASF is sufficiently practical as well.
Furthermore, the new definitions, linearity and orthogonality, high-order IIR and FIR realization of the PASF, and KF-PASF are expected to be the basis of future separation control studies on the quasi-periodic and quasi-aperiodic states.


\end{document}